\documentclass[acmsmall]{acmart}

\AtBeginDocument{%
  }

\usepackage[mode=build]{standalone}

\usepackage{graphicx}
\usepackage{url}
\usepackage{longtable}
\usepackage [capitalise]{cleveref}

\usepackage{dblfloatfix}

\usepackage{array}
\newcolumntype{x}[1]{>{\centering\arraybackslash\hspace{0pt}}p{#1}}

\newcolumntype{C}[1]{>{\centering\arraybackslash}p{#1}}

\usepackage{titlesec}

\titlespacing*{\subsection}
{0pt}{*2}{*1}

\usepackage{subcaption}

\usepackage{xcolor} 

\usepackage{tikz}

\usepackage{colortbl}

\usepackage{multirow}

\usepackage{pgfplots}

\newcommand{\blue}[1]{\textcolor{blue}{#1}}

\usepackage{xcolor}
\usepackage{hyperref}

\hypersetup{
  colorlinks=true,
  linkcolor=blue,
  citecolor=blue,
  filecolor=blue,
  urlcolor=blue,
}

\usepackage[nolist]{acronym}
\usepackage{mdframed}
\let\oldcite\cite
\renewcommand{\cite}[1]{\textcolor{black}{\oldcite{#1}}}

\usepackage{amsmath}

\usepackage{tcolorbox}

\usepackage{pifont}
\usepackage{float}
\usepackage{placeins}

\begin{document}

\noindent \textbf{Published version:}
This paper has been published in \textbf{ACM Computing Surveys}.
The final version is available at: https://dl.acm.org/doi/10.1145/3814955. \\





\title{Contextualizing Security and 
Privacy of Software-Defined Vehicles: A Literature Review and Industry Perspectives}


\author{Marco De Vincenzi}
\email{marco.devincenzi@iit.cnr.it}

\affiliation{%
  \institution{Institute for Informatics and Telematics, CNR}
  \streetaddress{Via Giuseppe Moruzzi 1}
  \city{Pisa}
  \state{}
  \country{Italy}
  \postcode{56124}
}

\author{Mert D. Pesé}
\email{mpese@clemson.edu}

\affiliation{%
  \institution{School of Computing, Clemson University}
  \streetaddress{215 McAdams Hall}
  \city{Clemson}
  \state{SC}
  \country{USA}
  \postcode{29634}
}

\author{Chiara Bodei}
\email{chiara.bodei@unipi.it}

\affiliation{%
  \institution{Department of Computer Science, Università di Pisa}
  \streetaddress{Largo Bruno Pontecorvo, 3}
  \city{Pisa}
  \state{}
  \country{Italy}
  \postcode{56127}
}

\author{Ilaria Matteucci}
\email{ilaria.matteucci@iit.cnr.it}
\affiliation{%
  \institution{Institute for Informatics and Telematics, CNR}
  \streetaddress{Via Giuseppe Moruzzi 1}
  \city{Pisa}
  \state{}
  \country{Italy}
  \postcode{56124}
}

\author{Richard R. Brooks}
\email{rrb@clemson.edu}

\affiliation{%
  \institution{Department of Electrical and Computer Engineering, Clemson University}
  \streetaddress{313C Riggs Hall}
  \city{Clemson}
  \state{SC}
  \country{USA}
  \postcode{29634}
}

\author{Monowar Hasan}
\email{monowar.hasan@wsu.edu}

\affiliation{%
  \institution{School of EECS, Washington State University}
  \streetaddress{355 NE Spokane St}
  \city{Pullman}
  \state{WA}
  \country{USA}
  \postcode{99164-2920}
}

\author{Andrea Saracino }
\email{andrea.saracino@santannapisa.it}

\affiliation{%
  \institution{TeCIP Institute, Sant'Anna School of Advanced Studies}
  \streetaddress{Piazza Martiri della Libertà 33}
  \city{Pisa}
  \state{}
  \country{Italy}
  \postcode{56127}
}
\author{Mohammad Hamad}
\email{mohammad.hamad@tum.de}

\affiliation{%
  \institution{Technical University of Munich}
  \streetaddress{Arcisstr., 21}
  \city{München}
  \state{}
  \country{Germany}
  \postcode{D-80333}
  }
  
\author{Sebastian Steinhorst}
\email{sebastian.steinhorst@tum.de}

\affiliation{%
  \institution{Technical University of Munich}
  \streetaddress{Arcisstr., 21}
  \city{München}
  \state{}
  \country{Germany}
  \postcode{D-80333}
}

 \renewcommand{\shortauthors}{De Vincenzi et al.}

\begin{abstract}
The growing reliance on software in road vehicles has led to the emergence of Software-Defined Vehicles (SDV). This work analyzes SDV security and privacy through a systematic literature review complemented by an industry questionnaire across the automotive supply chain. The analysis is structured as four research questions and results in a security framework serving as a roadmap for SDV protection. The findings emphasize addressing mixed-criticality architectural challenges, deploying layered security mechanisms, and integrating privacy-preserving techniques. 
The results highlight the need to harmonize in-vehicle and cloud-based defenses to strengthen cybersecurity and V2X resilience in Intelligent Transportation Systems (ITS).

\end{abstract}

\maketitle
\begin{acronym}
    \acro{ADAS}{Advanced Driver-Assistance Systems}
    \acro{AAOS}{Android Automotive OS}
    \acro{ARP}{Address Resolution Protocol}
    \acro{AI}{Artificial Intelligence}
    \acro{API}{Application Programming Interface}
    \acro{AV}{Autonomous Vehicle}
    \acro{CAN}{Controller Area Network}
    \acro{CI/CD}{Continuous Integration/Continuous Deployment}
    \acro{CV}{Connected Vehicle}
    \acro{DFS}{Depth-First Search}
    \acro{DL}{Deep Learning}
    \acro{DoS}{Denial-of-Service}
    \acro{E/E}{Electrical/Electronic}
    \acro{ECC}{Elliptic Curve Cryptography}
    \acro{ECU}{Electronic Control Unit}
    \acro{EV}{Electric Vehicle}
    \acro{GDPR}{General Data Protection Regulation}
    \acro{IDS}{Intrusion Detection System}
    \acro{IDPS}{Intrusion Detection and Prevention System}
    \acro{IMU}{Inertial Measurement Unit}
    \acro{IoT}{Internet of Things}
    \acro{IPS}{Intrusion Prevention System}
    \acro{ITS}{Intelligent Transportation System}
    \acro{IVI}{In-Vehicle Infotainment}
    \acro{IVN}{In-Vehicle Network}
    \acro{LR}{Literature Review}
    \acro{MAC}{Message Authentication Codes}
    \acro{ML}{Machine Learning}
    \acro{MitM}{Machine-in-the-Middle}
    \acro{MQTT}{Message Queuing Telemetry Transport}
    \acro{OBD-II}{On-Board Diagnostics-II}
    \acro{OEM}{Original Equipment Manufacturer}
    \acro{OTA}{Over-the-Air}
    \acro{HPC}{High-Performance Computing}
    \acro{PII}{Personally Identifiable Information}
    \acro{RQ}{Research Question}
    \acro{ROS}{Robot Operating System}
    \acro{SaaS}{Software as a Service}
    \acro{SVSE}{Secure Vehicle Software Engineering}
    \acro{SDN}{Software-Defined Networking}
    \acro{SDV}{Software-Defined Vehicle}
    \acro{SecOC}{Secure Onboard Communication}
    \acro{SCA}{Software Composition Analysis}
    \acro{SOA}{Service Oriented Architecture}
    \acro{S-ARP}{Secure ARP}
    \acro{SVSE}{Software Vehicle Security Engine}
    \acro{TAK}{Title-Abstract-Keywords}
    \acro{TARA}{Threat Analysis and Risk Assessment}
    \acro{TESLA}{Timed Efficient Stream Loss-Tolerant Authentication}
    \acro{TLS}{Transport Layer Security}
    \acro{ToS}{Terms of Service}
    \acro{V2I}{Vehicle-to-Infrastructure}
    \acro{V2X}{Vehicle-to-Everything}
    \acro{V2V}{Vehicle-to-Vehicle}
    \acro{VANET}{Vehicular Ad Hoc Networks}
    \acro{VLAN}{Virtual LAN}
    \acro{LLM}{Large Language Model}
\acro{VLM}{Vision-Language Model}
\acro{MLLM}{Multimodal Large Language Model}
\acro{VLA}{Vision-Language-Action model}
\acro{RAG}{Retrieval-Augmented Generation}
\acro{CLIP}{Contrastive Language--Image Pretraining}
\acro{SOA}{Service-Oriented Architecture}

\end{acronym}

\section{Introduction} \label{sec:introduction}

\acp{SDV} represent an emerging trend in the automotive industry \cite{s42154-022-00179-z, deloitteSoftwaredefinedVehicles, towardsautomotiveSoftwareDefined}, where vehicle functionalities, behaviors, and features are increasingly controlled by software rather than hardware components \cite{10394507,ibmSoftwareDefined, 10217971}. It is estimated that by 2040, 40\% of the total revenue of the automotive industry will come from digitally enabled services \cite{accenture2022software, hoyal2024software}. Recent industry analyses indicate that the convergence of \ac{AI} solutions, such as Vision-Language Models (VLM), and software-defined architectures is transforming vehicles into adaptive, continuously evolving platforms, positioning \acp{SDV} as a cornerstone for next-generation automotive innovation and competitiveness \cite{wardsauto2025_ai_sdv,mckinsey2025_software_defined_hardware_ai}. This paradigm shift suggests a gradual redesign of current architectures and the entire industry, providing users with benefits such as greater flexibility, easier upgrades, personalized customization, and the integration of autonomous technologies \cite{JamaWhitePaper}. This evolution affects not only development and operations, but also enables new business models and forms of collaboration, such as partnerships between \acp{OEM} and technology companies \cite{boschmobilityBoschSoftwaredefined}. In this context, firms traditionally rooted in the technology sector are increasingly entering the automotive domain and operating as vehicle manufacturers or mobility providers \cite{google_waymo2025, xiaomi_automotive2025}.

The growing integration of software components has been accompanied by a rise in cyberattacks. This trend dates back to one of the first remote vehicle exploits, which targeted a Jeep Cherokee in 2014 \cite{JeepAttack}. It continued in subsequent years, including the discovery of 14 vulnerabilities in BMW vehicles by engineers from the Keen Security Lab in 2018 \cite{tencent2018bmw}, and the identification of flaws in a Tesla Model 3 by a German teenager in 2022 \cite{cnnTeensTesla}. More recently, in 2024, Sam Curry and his team demonstrated the ability to remotely control Kia vehicle functionality using only the license plate \cite{curry2024hacking}.
Looking ahead, \ac{AI} is expected to become a key enabler for threat actors, allowing rapid vulnerability discovery and exploitation, and potentially supporting fleet-wide attacks \cite{upstreamReport}. Moreover, recent incidents involving modified communication devices that were remotely activated and controlled \cite{kingsley2024} have heightened concerns about the integrity of the production supply chain.
Such attacks, particularly when vehicles are deliberately deployed or compromised, may transform vehicles into weapons, as observed in vehicle ramming incidents, posing severe risks to public safety \cite{cisa2022vehicleramming, ramming1, ramming2, ramming3}. Finally, the increasing complexity and volume of vehicle software, together with the ongoing technological transformation of vehicles, have expanded the number of potential attack surfaces and vectors \cite{surfaces}.

From an industrial perspective, the ISO/SAE 21434 standard \cite{ISO21434} and the UNECE regulations R155/156 \cite{R155, R156} were introduced in 2021 to address cybersecurity vulnerabilities in \acp{CV} and \acp{SDV}. These security frameworks aim to harmonize and strengthen cybersecurity practices across the automotive industry.
However, as the transition toward UNECE R155 compliance advances, open questions remain regarding the implementation of new cybersecurity and software update infrastructures. A key challenge concerns the effective application of “Security by Design” principles in emerging software architectures. As an example, the de facto standard AUTomotive Open System ARchitecture (AUTOSAR) \cite{autosarAdaptivePlatform} introduced the Adaptive Platform to address the security requirements in \ac{SDV} architectures by adopting a dynamic, service-oriented, POSIX-based approach, in contrast to the Classic Platform tailored to hardware-centric vehicle architectures. This evolution is under active development as \ac{SDV} requirements continue to mature \cite{autosar_sdv_opening_strategy}.  Furthermore, additional challenges arise in coordinating cybersecurity with SOTIF (Safety of the Intended Functionality) and functional safety requirements defined by ISO 26262 \cite{ISO26262}.

This work investigates security and privacy issues in \acp{SDV} by analyzing insights from the literature and industry feedback on current challenges and potential solutions. As the concept of \ac{SDV} \cite{10217971} is still evolving and lacks a consolidated definition, foundational concepts relevant to \acp{SDV}' security are first introduced. These include a definition building on prior work \cite{10217971} and an analysis of the differences between \acp{SDV}, \acp{AV}, and \acp{CV}.
The study is conducted as a systematic literature review, structured around a foundational section and four \acp{RQ}. Relevant literature is systematically collected and analyzed following the review methodology described in \cref{sec:lr}. To further strengthen the analysis, feedback from industry experts is incorporated through a targeted elicitation (\cref{sec:questionnaire}). This integration addresses both the limited availability of literature on this emerging topic and the need to capture practical insights from professionals directly involved in \ac{SDV} development. Their contributions on key aspects, security threats, and mitigation strategies complement the literature and provide a more comprehensive view of the current state of \ac{SDV} development.

The results show that the transition to \acp{SDV} introduces challenges and risks in security and privacy, mainly due to the increased reliance on software and attack surfaces within vehicles. These risks range from API vulnerabilities and third-party software risks to complex supply chain threats and potential privacy infringements. Industry feedback further underscores the urgent need for robust, standardized security frameworks and privacy-preserving mechanisms to support the evolution of \acp{SDV}. In addition, the findings highlight the need to adopt multilayered security measures and to integrate in-vehicle and cloud-based solutions to protect future \acp{SDV}, while increasing user trust, particularly in the widespread adoption of \ac{AI}-based solutions within \acp{ITS}.

\subsection{Organization of the Paper}
This work is structured into three main sections: \textit{(I)} a foundational section (\cref{sec:sdvdefinition}) that presents a definition of \acp{SDV} to highlight their features, distinguish them as unique entities, and clarify the differences between \acp{SDV}, \acp{CV}, and \acp{AV} (\cref{sec:sdvdifferences}); \textit{(II)} a section (\cref{sec:attackSurfaces}) addressing \ac{SDV} security, which includes two research questions on attack surfaces (\cref{sec:lrrq1}, \textbf{\ac{RQ}1}) and mitigations (\cref{sec:lrrq2}, \textbf{\ac{RQ}2}); and \textit{(III)} a section (\cref{sec:securityChallenges}) on \ac{SDV} security challenges that examines two research questions focusing on \ac{OTA} (\cref{sec:lrrq3}, \textbf{\ac{RQ}3}) and \ac{SDV} data privacy (\cref{sec:lrrq4}, \textbf{\ac{RQ}4}). These aspects are selected for their critical roles in the functionality and security of \acp{SDV}. \ac{OTA} updates enable continuous improvement and maintenance of \acp{SDV}, while \acp{SDV} generate and process vast amounts of sensitive data. \Cref{fig:ourwork} summarizes the software-centric ecosystem, the proposed \acp{RQ}, and the main actors involved. It also illustrates the interconnections among \ac{SDV} security challenges, showing how they inform the threat and mitigation sections. Each section concludes with a gray box highlighting the key takeaways. Finally, \cref{sec:conclusion} presents the overall conclusions. The \acp{RQ} are as follows:

\noindent\textbf{\ac{RQ}1} - What are the attack surfaces and associated threats in \acp{SDV}?

\noindent\textbf{\ac{RQ}2} - What strategies can mitigate attack surface vulnerabilities?

\noindent\textbf{\ac{RQ}3} - What are the main security issues that \ac{OTA} updates face, including the outside vehicle environment?

\noindent\textbf{\ac{RQ}4} - How do \acp{SDV} affect data collection, and what are the primary concerns related to user and vehicle privacy?


\begin{figure}[t!]
    \centering
    \includegraphics[width=0.89\linewidth]{figures/schema.pdf}
    \caption{The review structure with the new software-centric ecosystem, including the \acp{RQ}.}
    \Description{The new Vehicular-Software (VehiSoft) centric ecosystem focuses on four key topics and the primary actors involved.}

    \label{fig:ourwork}
\end{figure}
\subsection{Motivations and Contributions}\label{sec:contribution}

Considering the rapid evolution of vehicles in recent years and the absence of a comprehensive definition and systematic analysis of \ac{SDV} security and privacy, the motivations behind this work are fourfold, as outlined below.

\noindent \textbf{Technological evolution:} the implications of the transition from hardware-centric to software-centric vehicle systems need to be explored, as this shift is rapidly redefining the boundaries and capabilities of modern vehicles \cite{10394507,ibmSoftwareDefined,10217971}.

\noindent \textbf{Security and privacy concerns:} the growing concerns surrounding security and privacy risks associated with \acp{SDV} need to be examined, particularly in light of recent high-profile cyberattacks on \acp{CV} and in the entire \ac{CV} ecosystem \cite{tencent2018bmw,cnnTeensTesla,curry2024hacking, 10818588}.

\noindent \textbf{Need for comprehensive analysis:} in view of the requirements set by UNECE regulations R155/156 \cite{R155,R156} and the ISO/SAE 21434 standard \cite{ISO21434}, a detailed review of existing literature and current practices is provided to identify knowledge gaps and potential areas for compliance and innovation in \acp{SDV} security and privacy \cite{10217971}.

\noindent \textbf{Supply chain complexity:} the \ac{SDV} supply chain comprises multiple layers of entities, including \acp{OEM} responsible for vehicle design and manufacturing, as well as suppliers across Tier~1, Tier~2, and Tier~3 levels. This layered structure increases complexity, as diverse software and hardware components from multiple suppliers must be integrated, often under heterogeneous cybersecurity standards \cite{keane2024software,boigon2024software}.
The analysis carried out in this work leads to the following contributions:

\noindent \textbf{Comprehensive review:} a complete review of the literature is presented, complemented by a questionnaire collecting insights from industry experts, to examine the current state of security and data privacy in \acp{SDV} while highlighting key challenges and emerging trends (\cref{sec:lr}).

\noindent \textbf{Definition of \acp{SDV}:} following \cite{10217971}, a definition of \acp{SDV} is provided to clearly distinguish among \acp{SDV}, \acp{AV}, and \acp{CV} (\cref{sec:sdvdefinition}).

\noindent \textbf{Attack surfaces and threats identification:} potential attack surfaces and threats specific to \acp{SDV} are identified and categorized, taking into account their unique characteristics and technological innovations (\cref{sec:lrrq1}).

\noindent \textbf{Mitigation strategies:} existing and emerging solutions for mitigating identified threats are examined (\cref{sec:lrrq2}), with a particular focus on \ac{OTA} updates (\cref{sec:lrrq3}) and data privacy (\cref{sec:lrrq4}).

\noindent \textbf{Takeaway and future directions:} open issues and future research directions are discussed, and a roadmap for enhancing the security and privacy of \acp{SDV} is outlined, with a dedicated summary box included in each section.

\section{Methodology} \label{sec:sources}
The following two subsections present the methodology used to retrieve evidence and data from the systematic literature review and the industry expert elicitation.

\subsection{Literature Review Process} \label{sec:lr}

The literature review for this study follows a systematic approach by adapting the SALSA (Search, Appraisal, Synthesis, and Analysis) method \cite{bookSystematicReview} to mine the relevant literature. The approach is based on the guidelines of Booth \textit{et al.} \cite{bookSystematicReview} and is tailored to meet the specific context of \acp{SDV}. The process begins with the formulation of \acp{RQ} and an exploration of background topics on \acp{SDV}. The output of this systematic process is a defined set of articles that address each RQ and topic. The workflow can be broken down into multiple stages of the method: Search, Appraisal, Synthesis, and Analysis. The search process is driven by the design of the \acp{RQ}. This is followed by the development of the query to gather relevant literature from the selected primary and secondary sources. 

Primary sources are databases that directly house academic research, while secondary sources provide aggregated or second-hand information and were used to supplement the search. The selected primary sources include relevant digital libraries such as IEEE Xplore \cite{IEEEXplore}, ACM Digital Library \cite{ACMDL}, Science Direct \cite{ScienceDirect}, and Springer Link \cite{Springer}. The selected secondary sources include Google Scholar \cite{GoogleScholar} and ResearchGate \cite{ResearchGate}. These secondary sources are known to return large volumes of results, and thus the search was limited to the first 100 results sorted by relevance. The query string ``\texttt{software AND defined AND vehicle(s)}'' was used 
across primary and secondary sources, with results filtered using \ac{TAK} criteria for relevance. After retrieving the articles, the next phase involved assessing them according to inclusion and exclusion criteria. Inclusion required articles to be in English, peer-reviewed, and focused on security and privacy in \acp{SDV}. Exclusion criteria removed publications before 2004 and unrelated topics such as ``Software-Defined Internet of Vehicles''. A snowballing phase using the Paperfetcher Tool \cite{Pallath2023-lm} followed, allowing backward and forward reference chasing to expand the selection. Due to the limited number of articles retrieved, additional tertiary sources were incorporated, including previous \ac{SDV} surveys and articles recommended by the authors. 

\begin{figure}[t!]
    \centering
    \includegraphics[width=\linewidth]{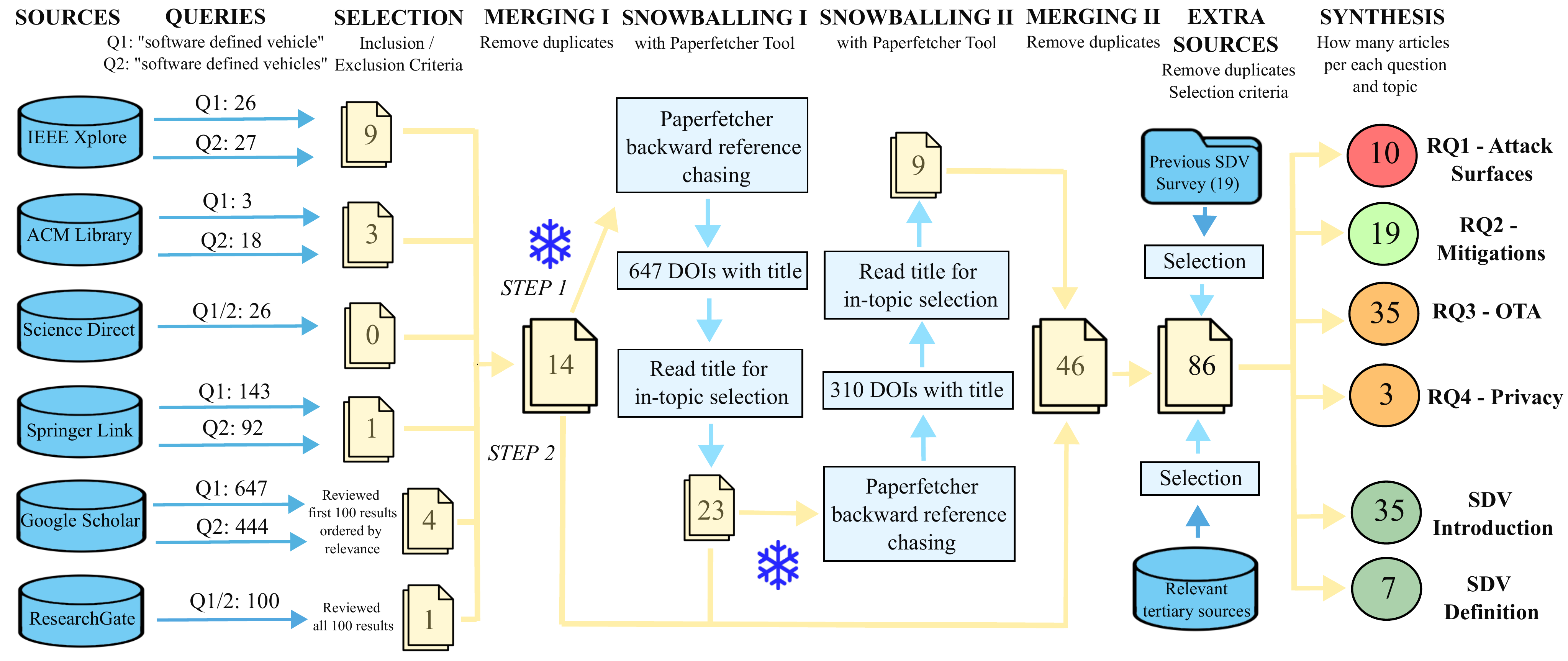}
    \caption{Literature review workflow schema.}
    \Description{A diagram showing the workflow schema of our search process.}
    \label{fig:workflowschema}
\end{figure}

In the synthesis phase, selected papers were categorized by formulated \acp{RQ} (\cref{fig:workflowschema}). The figure illustrates the workflow of the literature review with the steps described earlier. The numbers in the circles indicate the number of articles retrieved after each filtering step and, ultimately, assigned to each \ac{RQ} or topic. In the analysis phase, to answer the \acp{RQ}, at least three authors independently reviewed the articles in each category and formulated their assessments. These evaluations were then consolidated, and the corresponding sections were written by the designated lead author.

\subsection{Expert Elicitation}
\label{sec:questionnaire}

An expert elicitation methodology~\cite{colson2018expert} complements the literature review and mitigates blind spots arising from the emerging nature of \acp{SDV}, where evidence remains sparse for some aspects (e.g., RQ1 and RQ4). The elicitation was designed as an exploratory expert study based on purposive sampling and limited to \textit{industry participants}, aiming to capture high-signal practitioner judgments rather than population-level prevalence. To support replicability, the full questionnaire and anonymized responses are publicly available\footnote{https://github.com/Marc-cn/sdv-security-privacy-elicitation}, enabling reuse and validation. \cref{tab:questionnaire_characteristics} summarizes the structure of the elicitation.

\textit{Elicitation design}. The questionnaire was derived from the four research questions (\ac{RQ}1–\ac{RQ}4) and the foundational \ac{SDV} definition goals. As summarized in \cref{tab:questionnaire_characteristics}, it includes 65 items organized into five sections: (i) SDV definition, (ii) attack surfaces and threats, (iii) mitigation strategies, (iv) \ac{OTA} update risks, and (v) data privacy and governance. Most items used five-point Likert scales~\cite{likert1932technique} with section-specific anchors. Prior to deployment, the questionnaire underwent internal content review to ensure coverage of the \acp{RQ} and clarity.

\textit{Participant recruitment and composition}. Experts were recruited via purposive sampling from the \ac{SDV} supply chain, requiring professional involvement in automotive cybersecurity and exposure to at least one \ac{RQ}. A targeted list of 22 experts ensured coverage across roles and perspectives; participation was voluntary and anonymous. Eleven complete responses were collected (50\% response rate). Respondents represented \acp{OEM} (18.2\%), Tier~1 (27.3\%), Tier~2 (9.1\%), and other automotive cybersecurity stakeholders (45.5\%). Roles spanned engineering, technical leadership, senior expert, and executive positions.

\textit{Analysis methodology and metrics}. Analysis employs descriptive, consensus-oriented measures suitable for exploratory Likert-scale studies. For each item, responses are summarized using the median and interquartile range (IQR), defined as $\mathrm{IQR}=Q_3-Q_1$. The median reflects typical expert judgment, while the IQR captures agreement. For example, the criticality of ``Insecure \acp{API}'' (Q26) yielded a median of 5 and IQR of 1, indicating strong consensus. To complement these measures, Top-Two-Box Agreement (TTBA) is reported as the percentage of ratings equal to 4 or 5, providing an intuitive indicator of endorsement. Free-text responses were analyzed through lightweight thematic coding, revealing additional concerns such as fail-safe mechanisms and overlooked attacker profiles. As with all small-sample elicitations, results are indicative rather than statistically generalizable.

\textit{Presentation of elicitation results}. Elicitation outcomes are reported in dedicated ``Experts' feedback'' subsections.
Only representative items are reported and discussed, while all remaining items are made available online in the GitHub repository. Although limited in scope, combining expert input with existing research helps reveal both alignments and gaps between academic ideas and real-world perspectives, adding context to \ac{SDV} security challenges.

\begin{table*}[!t]
\centering
\footnotesize
\renewcommand{\arraystretch}{1}
\setlength{\tabcolsep}{6pt}
\caption{Structure and characteristics of the expert elicitation questionnaire.}
\label{tab:questionnaire_characteristics}
\begin{tabular}{p{1.8cm} p{3.8cm} p{1.4cm} p{2.6cm} p{2.2cm}}
\hline
\textbf{Section} & \textbf{Focus} & \textbf{\# Items} & \textbf{Item Type} & \textbf{Scale / Format} \\
\hline

Introduction &
Participant background (organization type, role) &
2 &
Categorical + free text &
Multiple choice; open text \\ \hline

SDV Definition &
Foundational SDV features and architectural principles &
19 (+1 open) &
Likert + open-ended &
1--5 (not relevant $\rightarrow$ highly relevant) \\ \hline

Security Threats &
Perceived criticality of SDV attack surfaces and risks &
12 &
Likert &
1--5 (low $\rightarrow$ high criticality) \\ \hline

Wireless Attack Vectors &
Criticality of wireless interfaces &
6 &
Likert (matrix) &
1--5 (not relevant $\rightarrow$ highly relevant) \\ \hline

Possible Attackers &
Relevance of attacker profiles &
4 (+2 open) &
Likert + open-ended &
1--5 (not relevant $\rightarrow$ highly relevant) \\ \hline

Security Mitigations &
Priority of technical, organizational, and regulatory mitigations &
14 (+1 open) &
Likert + open-ended &
1--5 (low $\rightarrow$ high priority) \\ \hline

OTA Security &
OTA risks, properties, and architectures &
3 (+1 open) &
Likert (grouped) + open-ended &
1--5 (low $\rightarrow$ high criticality) \\ \hline

Privacy &
Privacy governance, rights, and responsibilities in SDVs &
7 (+1 open) &
Likert + open-ended &
1--5 (strongly disagree $\rightarrow$ strongly agree) \\ \hline

\textbf{Overall} &
Expert elicitation across SDV lifecycle &
\textbf{65 total} &
Mixed (ordinal + qualitative) &
Anonymous, self-administered \\

\hline
\end{tabular}
\end{table*}

\section{Definition and Foundational Concepts of \ac{SDV}} \label{sec:sdvdefinition}

The concept of \acp{SDV} is relatively recent, and existing definitions and key elements vary across authors and industrial stakeholders, with limited research focusing explicitly on \acp{SDV}. While several works introduce broad definitions followed by specific viewpoints, industry-oriented white papers and reports tend to emphasize \ac{SDV} capabilities \cite{BoschBook,JamaWhitePaper}. As a result, a clear consensus on the most important \ac{SDV} features has yet to emerge.

To establish a common baseline, we adopt the definition provided by some of the authors in \cite{10217971}: \textit{``\ac{SDV} is an in-vehicle solution that enables abstraction and management of vehicle hardware components through software to create a scalable architecture with centralized and distributed local and remote controls. Furthermore, all vehicle software components must support \ac{OTA} updates''}. 
\acp{SDV} embody a software-centric paradigm in which vehicle functionality is continuously introduced and evolved through software, requiring new architectures to meet scalability demands while significantly expanding the attack surface and elevating security and privacy to core design concerns.

Starting from this conceptual foundation,
the main \ac{SDV} features are identified through a combination of literature analysis, domain knowledge, and contributions from a dedicated expert-based questionnaire. The experts reviewed the initial list and provided feedback, which was integrated into the final version. The resulting features are organized hierarchically and grouped into three main categories: (a) decoupling of hardware and software, (b) smart vehicle and paradigm shift, and (c) transformation of automotive architectures and enabling technologies. 
Related notions are then examined, and similarities between \acp{SDV} and other devices or technologies are highlighted to further define and contextualize the \ac{SDV} concept.

\subsection{Core Features of Software-Defined Vehicles}
This section details the core features that characterize \acp{SDV}, expanding on the three main categories introduced above. These features capture the architectural, functional, and systemic changes enabled by \acp{SDV} and form the basis for the subsequent analysis of security and privacy challenges.

\paragraph{\bf (a) Decoupling of Hardware and Software.} 

It represents a fundamental shift in vehicle design, where software increasingly defines functionality and hardware assumes an abstracted supporting role, enabling flexibility, scalability, and continuous evolution compared to traditional hardware-centric models \cite{s42154-022-00179-z, Shirasat24,10.1109/MetroCAD56305.2022.00015}. In this paradigm, vehicles are primarily software-defined, with hardware providing execution and sensing, enabling modularity, upgradeability, dynamic resource reassignment, and runtime deployment or migration of services without hardware changes.
The following list summarizes the main related features identified in the literature.

\textit{Hardware becomes an abstract shared resource}: Software can be called/accessed, allowing a flexible combination that enhances both functionality and performance \cite{s42154-022-00179-z}. Furthermore, hardware abstraction plays a crucial role in effectively managing underlying hardware functions and application services such as described in the AUTOSAR Adaptive Platform with its \ac{SOA} \cite{autosarAdaptivePlatform}.
This can be accomplished through the use of standardized interfaces to abstract hardware specifics, enabling developers to design software and applications without having to adapt to the unique characteristics of each hardware model.
\cite{s42154-022-00179-z,10.1016/j.comcom.2018.09.010}. 

\textit{Hardware and Software agnosticism}: Ideally, software services and interfaces (such as \acp{API} and middleware) should be hardware-agnostic to allow interaction of applications with other hardware services or communication with other applications \cite{Shirasat24}. This approach enables the sharing of solutions among manufacturers, suppliers, and academia. Software applications can be developed independently and tailored to specific functions or services and not to vehicle types \cite{Becker22}. 

\textit{Hardware/Software updateability}: Vehicle functions and capabilities, powered by software, can be 
upgraded and managed continuously throughout the vehicle's entire lifecycle. 
These updates can also occur after sale \cite{s42154-022-00179-z,Shirasat24}.
All vehicle components must support continuous \ac{OTA} software updates, allowing for the agile \cite{Shirasat24} addition of new features, fixing vulnerabilities or software bugs, and optimizing existing functions \cite{s42154-022-00179-z,Becker22,10.1109/MetroCAD56305.2022.00015}.
This adaptability also enables vehicles to stay current and extend their useful life. As vehicle technology becomes increasingly software-centric, the reliance on software introduces challenges similar to those faced by computers and smartphones, such as software obsolescence, hardware compatibility issues, cybersecurity risks, and higher maintenance costs.

\textit{Enhancement of Software reuse}: Reusing software components across different systems and platforms enhances development efficiency, reduces costs, and can ensure quality consistency \cite{10.1016/j.comcom.2018.09.010}.

\paragraph{\bf (b) Smart Vehicle and Paradigm Shift.} 

\acp{SDV} enable a paradigm shift in which vehicles become smart, data-centric systems capable of learning, adaptation, and continuous improvement. Through extensive sensing, connectivity, and computation, \acp{SDV} unlock data value across the vehicle lifecycle and integrate vehicles into a broader digital ecosystem.
With increased integration of sensors, devices, and computing and communication capabilities, \acp{SDV} can manage and enhance data quality through closed-loop systems, enabling onboard software to support self-learning and evolution using \ac{AI}. This development increases personalization of user experiences and supports large-scale data processing by connecting vehicles to cloud platforms \cite{s42154-022-00179-z}. Moreover, this shift positions the vehicle as part of a broader ecosystem, where customer-centric development continuously refines features based on real-world usage \cite{Becker22}.


\paragraph{\bf (c)
Transformation of Automotive Architectures and Enabling Technologies.}

The transition toward \acp{SDV} is not limited to vehicle functionality, but is driving a broader transformation of the automotive industry. 
Traditional vehicle designs and development models are no longer sufficient to support \ac{SDV} requirements. In this context, architectural and connectivity aspects emerge as key enablers of \acp{SDV}. In particular, the
increasing complexity and volume of software require greater computing power 
\cite{s42154-022-00179-z} and
increased security measures \cite{Becker22}. Hardware, software, and communication architectures must evolve accordingly.

\textit{Architecture}: \acp{SDV} do not mandate a specific underlying architecture; however, zonal architectures align well with \ac{SDV} objectives, driving a shift from domain-centric to zonal-centric E/E designs \cite{10.1109, 10.1109/MetroCAD56305.2022.00015}. Thus, systems are organized by physical location rather than function or domain. Each zone is managed by a dedicated controller, with zonal controllers interconnected and communicating with a central computing system. Additionally, the vehicle architecture must support scalable resources to accommodate the requirements of \ac{V2X} communications.

\textit{On-board operating system}: This is a strategic component that enhances product competitiveness by providing integration and flexibility for vehicle control and application services. 
This system will evolve from partially integrated systems to comprehensive vehicle-level systems that support rich application ecosystems, improve control flexibility, and reduce integration costs \cite{s42154-022-00179-z}.

\textit{Automotive Ethernet}: 
It seems to be a good candidate for backbone communications in \ac{SDV} architecture where a large amount of data is transmitted \cite{10.1145/3637059}. 

\textit{Cloud-native design paradigm}: These paradigms should be adopted across hardware platforms from the cloud to the
vehicle-edge, following an automotive DevOps (Development Operations) perspective \cite{Shirasat24}.

Taken together, these core functionalities transform the vehicle into a continuously evolving software platform rather than a static product. While this shift enables unprecedented flexibility and innovation, it also requires rethinking security, safety, and governance across the entire lifecycle.

To conclude, the transition toward \acp{SDV} is often compared to the evolution of smartphones, reflecting a shift toward software-defined functionality and continuous feature updates \cite{s42154-022-00179-z,Becker22,Shirasat24}. While this analogy reflects the increasing role of software, abstraction, and \ac{OTA} updates, automotive systems face challenges that differ from those of consumer electronics. In particular, \acp{SDV} must handle mixed-criticality workloads that combine safety and real-time constraints with high computational demands. They also resemble robotic systems in their reliance on frameworks, libraries, and \acp{API} to manage complex sensor–actuator networks. A representative example is Apex.OS, a \ac{ROS}~2 variant developed by Apex.AI, which adapts robotics-oriented architectures to automotive systems \cite{Becker22}. Finally, \acp{SDV} adopt concepts from \ac{SDN}: by decoupling network control from data forwarding, \ac{SDN} principles enable centralized management, efficient communication, and dynamic resource allocation, supporting real-time coordination and cloud-based interaction \cite{renesas2024sdn}.

\subsection{Experts’ Feedback}


Focusing on the most relevant items from the elicitation (with full results available in the GitHub repository, as noted in Section~\ref{sec:questionnaire}), the highest level of expert consensus was observed for the decoupling of hardware and software, which received a median rating of 5 with low dispersion (IQR = 1) and a Top-Two-Box Agreement (TTBA) of 90.9\%. These values indicate both strong endorsement and tight agreement among experts on its fundamental role in enabling flexibility, scalability, and modularity in \acp{SDV}. This result aligns with the emphasis found in the literature and reinforces its characterization as a core SDV feature. Other architectural enablers also exhibit high levels of endorsement. Automotive Ethernet received a median rating of 4, with low dispersion (IQR = 1) and a TTBA of approximately 72.7\%, reflecting broad recognition of its importance as a communication backbone for SDVs. Similarly, continuous \ac{OTA} updates and advanced data management were rated highly (median = 4--5, IQR = 1), with TTBA values above 80\%, highlighting their perceived relevance for maintaining vehicle functionality and supporting long-term software evolution. In contrast, features related to generic software platforms show greater dispersion, with a higher interquartile range (IQR = 2) and a lower TTBA (approximately 63.6\%). This pattern suggests that while cross-platform and reusable software is widely acknowledged as relevant, experts differ more strongly in how they prioritize it relative to foundational architectural features. 

\definecolor{LightGreen}{RGB}{160,193,90}    
\definecolor{LightYellowGreen}{RGB}{173,214,51} 
\definecolor{BrightYellow}{RGB}{255,217,52}  
\definecolor{Orange}{RGB}{255,178,52}        
\definecolor{LightRed}{RGB}{255,140,90}      

\begin{figure}[t!]
\centering
\label{fig:definition}
\scalebox{0.80}{
\begin{tikzpicture}
  \begin{axis}[
    xbar stacked,
    width=0.7\textwidth,
    height=2cm,
    bar width=15pt,
    enlargelimits=0.1,
    legend style={
       at={(1.02,0.5)}, 
  anchor=west,
  draw=none
    },
    xlabel={Percentage of participants (\%)},
    symbolic y coords={
      Advanced data management,
      Continuous OTA updates,
      Automotive Ethernet,
      Generic Software,
      Decoupling Hardware and Software
    },
    ytick=data,
    y tick label style={align=right, anchor=east},
    y=0.7cm,
    nodes near coords,
    point meta=explicit symbolic,
    every node near coord/.append style={
      font=\small,
      align=center,
      inner sep=1pt,
      yshift=-3pt,
      xshift=-9pt,
      text=black
    },
    grid=none,
    xmajorgrids=true,
    ymajorgrids=false,
    grid style={dashed},
    xtick={0,10,...,100},
    xmin=0, xmax=100,
  ]

  \addplot+[xbar, fill=LightRed, draw=black] plot coordinates {
    (0,Advanced data management)[]
    (0,Continuous OTA updates)[]
    (0,Automotive Ethernet)[]
    (0,Generic Software)[]
    (0,Decoupling Hardware and Software)[]
  };

  \addplot+[xbar, fill=Orange, draw=black] plot coordinates {
    (9.1,Advanced data management)[9.1]
    (0,Continuous OTA updates)[]
    (9.1,Automotive Ethernet)[9.1]
    (0,Generic Software)[]
    (9.1,Decoupling Hardware and Software)[9.1]
  };

  \addplot+[xbar, fill=BrightYellow, draw=black] plot coordinates {
    (9.1,Advanced data management)[9.1]
    (0,Continuous OTA updates)[]
    (0,Automotive Ethernet)[]
    (27.3,Generic Software)[27.3]
    (0,Decoupling Hardware and Software)[]
  };

  \addplot+[xbar, fill=LightYellowGreen, draw=black] plot coordinates {
    (63.6,Advanced data management)[63.6]
    (45.5,Continuous OTA updates)[45.5]
    (45.5,Automotive Ethernet)[45.5]
    (36.4,Generic Software)[36.4]
    (27.3,Decoupling Hardware and Software)[27.3]
  };

  \addplot+[xbar, fill=LightGreen, draw=black] plot coordinates {
    (18.2,Advanced data management)[18.2]
    (54.5,Continuous OTA updates)[54.5]
    (45.5,Automotive Ethernet)[45.5]
    (36.4,Generic Software)[36.4]
    (63.6,Decoupling Hardware and Software)[63.6]
  };

  \legend{
    \strut Irrelevant,
    \strut Slightly,
    \strut Moderately,
    \strut Strongly,
    \strut Highly Relevant
  }

  \end{axis}
\end{tikzpicture}
}
\caption{Experts' answers for the definition section.}
\label{fig:definition_fig}
\Description{Questionnaire definition results.}
\end{figure}
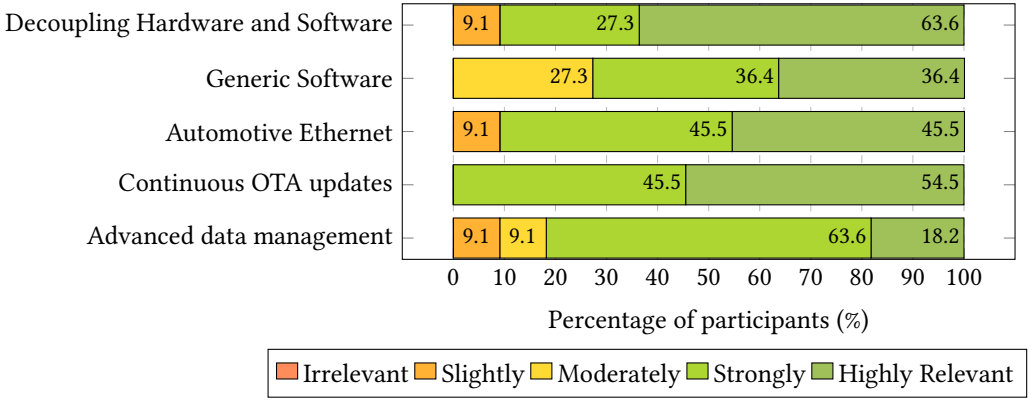

\begin{tcolorbox}[colback=gray!10, colframe=black, boxrule=0.3pt,left=2.5pt,right=2.6pt,top=1.5pt,bottom=1.5pt]
\textbf{Takeaway 1. } 
The concept of \acp{SDV} is still evolving. This analysis reveals variations in the definitions of key elements across authors and industrial stakeholders, as well as a limited body of research explicitly dedicated to \acp{SDV}. As a result, a set of core \ac{SDV} features is identified and their relevance is discussed based on both the literature and expert insights. This characterization provides a foundation for examining the associated security and privacy implications, while also offering guidance on aspects of \acp{SDV} that are likely to gain importance in the near future.
\end{tcolorbox}

\subsection{Differences between \ac{AV}/\ac{CV}/\ac{SDV}} \label{sec:sdvdifferences}
Although these three terms are frequently used interchangeably, \acp{AV}, \acp{CV}, and \acp{SDV} refer to distinct, yet partially overlapping, categories of vehicles. 

\textit{\ac{AV}} refers to vehicles capable, at different levels, of driving without human intervention, regardless of whether their design includes controls for a driver \cite{nhtsaAutomatedVehicles}. 
\acp{AV} may operate across different levels of driving automation, as defined by the SAE J3016 standard, ranging from Level~0 (no automation) and driver assistance (Levels~1--2) to conditional, high, and full automation (Levels~3--5), with the ability to transition between autonomous and human-controlled operation depending on system capabilities and operational conditions \cite{ZHAO2024122836,SAEJ3016}.
These vehicles use cameras and sensors to monitor the road and surroundings 
and make driving decisions, leveraging
artificial intelligence to handle driving tasks, manage traffic scenarios, and prevent accidents. \acp{AV} may also communicate with other
vehicles,
using visual signals such as blinkers, 
and dedicated protocols to coordinate maneuvers safely \cite{7535434}.

\textit{\acp{CV}} are vehicles equipped with \ac{V2X} communication technology, 
enabling interaction with other vehicles, infrastructure, and road users through 
back-end systems that facilitate data exchange \cite{uscv}.
This includes \ac{V2V} and \ac{V2I} communication, which use dedicated short-range radio signals to share information on vehicle speed, position, and direction, as well as road conditions and other relevant data \cite{uscv}. 
The concept of connected vehicles dates back to the mid-1990s, and many contemporary vehicles already fall into this category. Connectivity can be considered an enabling factor for both \ac{AV} and \ac{SDV} functionalities. 

\textit{\ac{SDV}} is the most recent of these three concepts and, as described in this section, refers to vehicles in which
hardware components and functionalities are managed by software \cite{10217971}. 
Beyond architectural implications already described, \acp{SDV} also enable new business models, such as subscription-based access to vehicle functionalities \cite{idtechex_sdv_business_models}.
Depending on the features that the driver intends to use, 
specific vehicle functionalities can be enabled for
a predetermined fee. 
This model allows users to pay only for the features they actually need at a given time. For instance, during extended periods of vehicle inactivity, users may temporarily disable advanced autonomous driving or comfort features, thereby avoiding subscription costs when such functionalities are not required. 
In \acp{SDV},
enabled and disabled features 
can range from an advanced navigation system to Level 3 autonomous driving and 
even performance-related settings such as speed or torque limiters tailored to different driving styles.
Since these features are enabled or disabled through  \ac{OTA} configurations, \acp{SDV} 
typically require continuous network connectivity and therefore often also qualify as \acp{CV}.

\acp{SDV} differ from \acp{AV} and \acp{CV} by adopting a software-centric control and evolution model in which vehicle functions are centrally orchestrated and continuously upgradable, enabling faster innovation and new maintenance and business models compared to function-bound ECUs. This paradigm naturally underpins scalable connectivity and higher levels of automation by supporting coordinated software control, data fusion, and cross-domain integration.

In summary, \ac{SDV}, \ac{AV}, and \ac{CV} represent distinct yet overlapping characteristics of modern vehicles. Most vehicles today qualify as \acp{CV}, while those equipped with varying levels of \ac{ADAS} or autonomous driving features fall under the category of \acp{AV}.
 On the other hand, \acp{SDV} blur the lines by integrating software-managed features, potentially including autonomous driving as a configurable option. While a vehicle can be connected 
without being autonomous (\ac{CV} only), or autonomous without being connected (\ac{AV} only), such configurations are increasingly uncommon and we do not have clear evidence of this choice. 
However, \ac{SDV} often require connectivity, though some configurations could be made through physical connections to the vehicle. This interplay highlights the nuanced relationships among these categories and 
reflects the ongoing evolution of vehicle technologies.

\section{SDV Attack Surfaces and Mitigations} \label{sec:attackSurfaces}
In this section, the security of \ac{SDV} is discussed by addressing two interconnected research questions: \textbf{\ac{RQ}1} focuses on identifying attack surfaces and threats (\cref{sec:lrrq1}), while \textbf{\ac{RQ}2} examines how the discovered threats can be mitigated (\cref{sec:lrrq2}), with their interrelation discussed in \cref{sec:discussionVulnMit}.

\subsection{\ac{RQ}1 - What are the attack surfaces and associated threats in \acp{SDV}?}
\label{sec:lrrq1}

\begin{figure}[t!]
	\centering
	\includegraphics[width=0.70\linewidth]{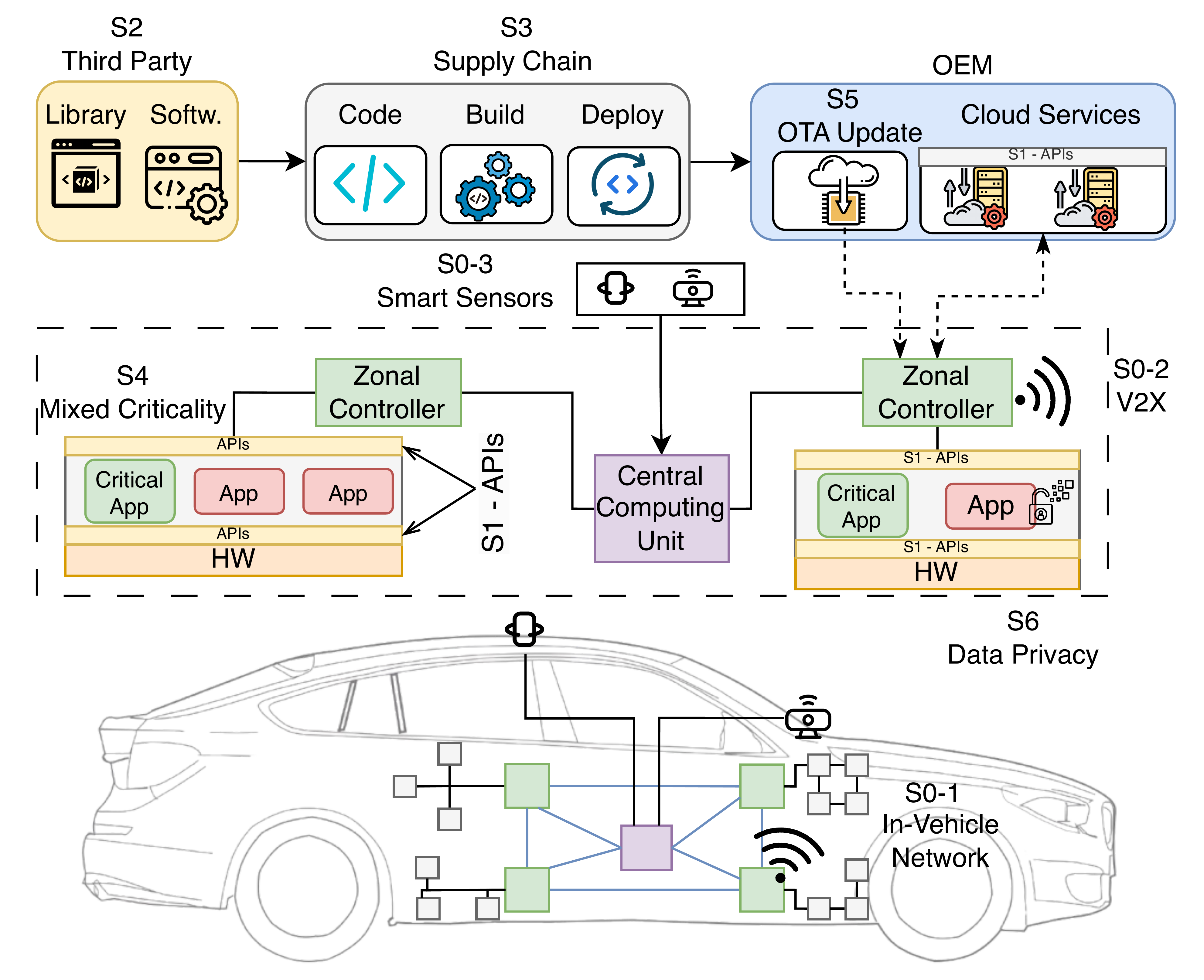}
	\caption{ Attack surfaces taxonomy, including both legacy (S0-1, S0-2, S0-3) and \ac{SDV} attack surfaces (S1-S6). }
	\label{fig:attackSurfaces}
	\Description{\ac{SDV} possible attack surfaces.}
\end{figure}

\acp{SDV} expose multiple attack surfaces that span software, communication, and data-management layers. Each identified attack surface is assigned a unique identifier (S1, S2, …) to allow concise referencing throughout the text and to link it explicitly to the corresponding threats (T1, T2, …) and mitigations discussed in \cref{sec:lrrq2}.
It is essential to note that the focus is limited to attack surfaces related to the specific characteristics of \acp{SDV}. Legacy attack surfaces inherited from autonomous and connected vehicles are nonetheless acknowledged, including S0-1 in-vehicle networks (e.g., \ac{CAN}) \cite{avatefipour2018stateoftheartsurveyinvehiclenetwork, Bodei2024CINNAMON}, Automotive Ethernet \cite{10.1145/3431233, 10.1145/3637059}, S0-2 \ac{V2X} communication \cite{ghosal2020security}, and S0-3 smart sensors \cite{9519442}. These surfaces are still relevant to \acp{SDV} and must be considered. \cref{fig:attackSurfaces} summarizes the possible attack surfaces and includes several layers of interaction. Critical components such as zonal controllers, central computing units, and external third-party libraries are highlighted as key vulnerabilities, exposing the \ac{SDV} to various cybersecurity risks from the supply chain to \ac{OTA} updates.
Based on the literature review, the following attack surfaces and related threats should be considered.
They are interpreted in light of the defining characteristics of \acp{SDV} which contribute to multiple and sometimes overlapping points of exposure.

\noindent\textbf{S1 - Insecure \acp{API}}: \acp{API} in \acp{SDV} enable interaction between software components across in-vehicle and backend systems, but the exposure significantly enlarges the attack surface. Misconfigured \acp{API} can allow authentication bypass, unintended access to production endpoints from development environments, and the invocation of backend services without valid access tokens \cite{eaton2023toyota}. Hard-coded environment configurations and client-side logic enabled attackers to switch from development to production \acp{API}, bypass corporate login mechanisms, and issue unauthenticated requests. This resulted in large-scale exposure of sensitive customer data, demonstrating how insecure \ac{API} design can directly lead to data exfiltration and privacy violations. As a recent real-world example, the 2024 Kia connected-vehicle breach demonstrated how authentication bypass in exposed backend \acp{API} enabled unauthorized access to vehicle functions and sensitive customer data \cite{curry2024hacking}. \noindent\textit{Main Threats:} client-side authentication bypass (\textbf{T1}) \cite{apisecurity_issue272}; unauthorized production \acp{API} access (\textbf{T2}); sensitive customer data exfiltration (\textbf{T7}) \cite{eaton2023toyota}.

\noindent\textbf{S2 - Third-party Applications and Libraries}: 
\acp{SDV} increasingly depend on third-party applications, frameworks, and software libraries to accelerate development, promote reuse, and support rapid deployment. While this improves modularity and reduces time-to-market, it introduces a significant attack surface, as externally sourced components may contain vulnerabilities or malicious functionality. Evidence from adjacent domains, including mobile platforms \cite{poeplau2014execute,enck2011study} and industrial control systems \cite{CVE-2023-1709}, shows that third-party software can be exploited to inject malicious code, bypass security mechanisms, or exfiltrate sensitive data. In the automotive context, inadequate vetting, missing integrity verification, or insecure update mechanisms for third-party components can enable unauthorized access to vehicle subsystems, exposure of personal and vehicular data (e.g., location and usage patterns), or interference with core vehicle functions, ultimately affecting safety and reliability \cite{hsu2014toyota}.  
\noindent\textit{Main Threats:} dynamic malicious code injection (\textbf{T3}) \cite{poeplau2014execute}; distribution of malicious framework updates \cite{poeplau2014execute} (\textbf{T4}); sensitive customer data exfiltration (\textbf{T7})  \cite{enck2011study}.

\noindent\textbf{S3 - Supply-Chain Security}: 
Unlike S2, which focuses on runtime third-party software dependencies, this attack surface is broader and concerns the integrity of the entire production and distribution pipeline. \ac{SDV} ecosystems depend on complex global supply chains spanning hardware manufacturing, firmware development, and software integration, introducing multiple points of trust violation. Prior work highlights how adversaries can exploit these dependencies to inject malicious artifacts into otherwise legitimate components, either by introducing counterfeit hardware \cite{guin2014counterfeit} or by compromising build, packaging, signing, or distribution workflows \cite{10179304}. In automotive systems, compromised supply-chain elements can undermine both safety and security guarantees, as malicious modifications may remain dormant until deployment or activation under specific conditions. 
\noindent\textit{Main Threats:} distribution of tampered firmware updates or compromised software build (\textbf{T4}) \cite{10179304}; counterfeit and recycled hardware components (\textbf{T10}) \cite{guin2014counterfeit}.

\noindent\textbf{S4 - Mixed Criticality}: 
\acp{SDV} increasingly adopt a \ac{SOA} to support dynamic service deployment and flexible interaction among software components with heterogeneous criticality levels \cite{rumez2020overview, LuoSembera2023VicOne}. While this paradigm improves modularity and updatability, it weakens the strict isolation traditionally enforced in safety-critical automotive architectures. Prior work shows that shared communication channels, middleware, and computational resources can cause unintended interference between services of different criticality \cite{lee2024dynamic}. In particular, unregulated service discovery, runtime binding, and resource contention may allow low-criticality services to affect the availability or integrity of high-criticality functions, potentially violating real-time and safety guarantees. These risks are concentrated at zonal controllers and central compute units, which host or coordinate multiple services and become high-value targets. 
\noindent\textit{Main Threats:} unauthorized service interaction and access (\textbf{T2})  \cite{rumez2020overview}; service-level \ac{DoS} (\textbf{T5}) \cite{rumez2020overview}; cross-criticality resource interference (\textbf{T6}) \cite{lee2024dynamic}.

\noindent\textbf{S5 - \ac{OTA} Update}: 
\ac{OTA} updates enable remote software deployment across vehicle components, reducing maintenance costs and recall overhead. However, the OTA channel constitutes a safety-critical attack surface, as it requires privileged access to in-vehicle networks and ECUs \cite{halder2020secure}. Weak protection of update authenticity, integrity, or confidentiality can allow adversaries to inject or modify firmware during transmission or installation. Recent work demonstrates that compromised firmware can be customized and installed remotely, enabling persistent control over infotainment systems \cite{DBLP:journals/virology/CostantinoVM24}. In addition, compromised update servers or distribution infrastructures may propagate malicious updates at scale, while insufficient version control exposes vehicles to downgrade and rollback attacks that reintroduce known vulnerabilities \cite{asokan2018assured}. \textbf{\ac{RQ}3} (Section~\ref{sec:lrrq3}) provides a detailed discussion of this topic. \noindent\textit{Main Threats:} authentication bypass (\textbf{T1}) \cite{curry2024hacking}; malicious firmware injection via OTA (\textbf{T3}) \cite{halder2020secure}; large-scale distribution of compromised updates (\textbf{T4}) \cite{DBLP:journals/virology/CostantinoVM24}; rollback to vulnerable firmware versions (\textbf{T9}) \cite{asokan2018assured}.

\noindent\textbf{S6 - Data Privacy}:
With the advent of \acp{SDV}, large volumes of data are continuously collected through onboard sensors and telematics units for diagnostics, service provisioning, and monetization by \acp{OEM} and third-party entities. As connectivity and data sharing increase, privacy risks intensify, since vehicular data often contains \ac{PII} \cite{gazdag2023privacy,pese2019survey}, including precise geolocation and fine-grained behavioral information. Privacy attacks primarily target the misuse or extraction of such data by (i) unauthorized external parties or (ii) authorized entities exceeding their intended access scope or purpose \cite{pese2023pricar}. These threats challenge consent, transparency, and regulatory compliance, particularly under data protection frameworks. \textbf{\ac{RQ}4} (Section~\ref{sec:lrrq4}) provides a detailed discussion of this topic. \noindent\textit{Main Threats:} unauthorized secondary use of personal data \cite{pese2023pricar} (\textbf{T7});  location and trajectory inference attacks \cite{gazdag2023privacy}  and driver re-identification from vehicular data (\textbf{T8}) \cite{gazdag2023privacy,pese2019survey}.

\subsubsection{\textbf{Experts' Feedback}} 

\definecolor{LightGreen}{RGB}{160,193,90}        
\definecolor{LightYellowGreen}{RGB}{173,214,51} 
\definecolor{BrightYellow}{RGB}{255,217,52}     
\definecolor{Orange}{RGB}{255,178,52}           
\definecolor{LightRed}{RGB}{255,140,90}         

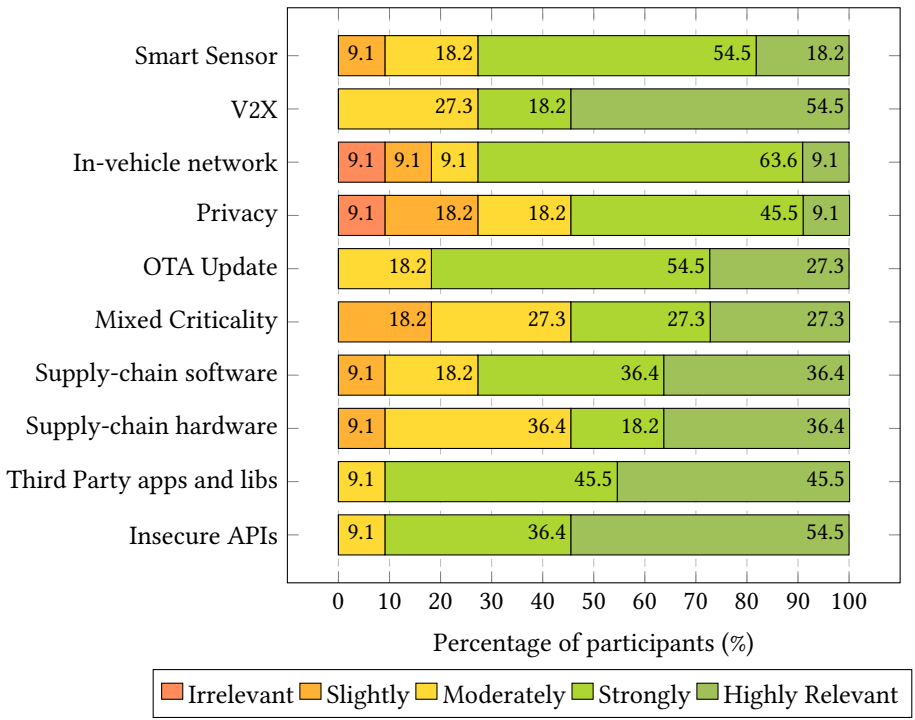
\begin{figure}[t!]
\centering
\label{fig:RQ1_graph}
\scalebox{0.80}{
\begin{tikzpicture}
  \begin{axis}[
    xbar stacked,
    width=0.7\textwidth,
    height=2cm,
    bar width=15pt,
    enlargelimits=0.1,
    legend style={
       at={(1.02,0.5)}, 
  anchor=west,
  draw=none
    },
    xlabel={Percentage of participants (\%)},
    symbolic y coords={
      Insecure APIs,
      Third Party apps and libs,
      Supply-chain hardware,
      Supply-chain software,
      Mixed Criticality,
      OTA Update,
      Privacy,
      In-vehicle network,
      V2X,
      Smart Sensor
    },
    ytick=data,
    y tick label style={align=right, anchor=east},
    y=0.7cm,
    nodes near coords,
    point meta=explicit symbolic,
    every node near coord/.append style={
      font=\small,
      align=center,
      inner sep=1pt,
      yshift=-3pt,
      xshift=-9pt,
      text=black
    },
    grid=none,
    xmajorgrids=true,
    ymajorgrids=false,
    grid style={dashed},
    xtick={0,10,...,100},
    xmin=0, xmax=100,
  ]

  \addplot+[xbar, fill=LightRed, draw=black] plot coordinates {
    (0,Insecure APIs)[] (0,Third Party apps and libs)[]
    (0,Supply-chain hardware)[] (0,Supply-chain software)[]
    (0,Mixed Criticality)[] (0,OTA Update)[]
    (9.1,Privacy)[9.1] (9.1,In-vehicle network)[9.1]
    (0,V2X)[] (0,Smart Sensor)[]
  };

  \addplot+[xbar, fill=Orange, draw=black] plot coordinates {
    (0,Insecure APIs)[] (0,Third Party apps and libs)[]
    (9.1,Supply-chain hardware)[9.1] (9.1,Supply-chain software)[9.1]
    (18.2,Mixed Criticality)[18.2] (0,OTA Update)[]
    (18.2,Privacy)[18.2] (9.1,In-vehicle network)[9.1]
    (0,V2X)[] (9.1,Smart Sensor)[9.1]
  };

  \addplot+[xbar, fill=BrightYellow, draw=black] plot coordinates {
    (9.1,Insecure APIs)[9.1] (9.1,Third Party apps and libs)[9.1]
    (36.4,Supply-chain hardware)[36.4] (18.2,Supply-chain software)[18.2]
    (27.3,Mixed Criticality)[27.3] (18.2,OTA Update)[18.2]
    (18.2,Privacy)[18.2] (9.1,In-vehicle network)[9.1]
    (27.3,V2X)[27.3] (18.2,Smart Sensor)[18.2]
  };

  \addplot+[xbar, fill=LightYellowGreen, draw=black] plot coordinates {
    (36.4,Insecure APIs)[36.4] (45.5,Third Party apps and libs)[45.5]
    (18.2,Supply-chain hardware)[18.2] (36.4,Supply-chain software)[36.4]
    (27.3,Mixed Criticality)[27.3] (54.5,OTA Update)[54.5]
    (45.5,Privacy)[45.5] (63.6,In-vehicle network)[63.6]
    (18.2,V2X)[18.2] (54.5,Smart Sensor)[54.5]
  };

  \addplot+[xbar, fill=LightGreen, draw=black] plot coordinates {
    (54.5,Insecure APIs)[54.5] (45.5,Third Party apps and libs)[45.5]
    (36.4,Supply-chain hardware)[36.4] (36.4,Supply-chain software)[36.4]
    (27.3,Mixed Criticality)[27.3] (27.3,OTA Update)[27.3]
    (9.1,Privacy)[9.1] (9.1,In-vehicle network)[9.1]
    (54.5,V2X)[54.5] (18.2,Smart Sensor)[18.2]
  };

  \legend{
    \strut Irrelevant,
    \strut Slightly,
    \strut Moderately,
    \strut Strongly,
    \strut Highly Relevant
  }

  \end{axis}
\end{tikzpicture}
}
\caption{The experts' answers to \ac{RQ}1, where a rating of 1 indicates irrelevance and 5 indicates high relevance for \ac{SDV} security.}
\label{fig:RQ1_answers}
\Description{Questionnaire RQ1 results.}
\end{figure}

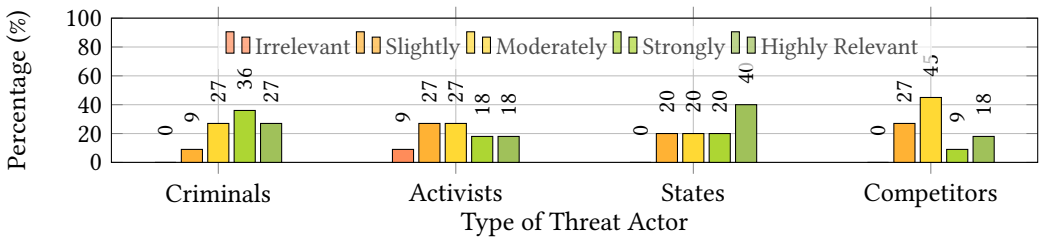
\begin{figure}[t!]
	\centering
	\scalebox{0.89}{ 
	\begin{tikzpicture}
		\begin{axis}[
			ybar,
			ymin=0, ymax=100,
			width=\textwidth,
			height=0.25\textwidth,
			bar width=8pt,
			xlabel={Type of Threat Actor},
			ylabel={Percentage (\%)},
			xtick=data,
			symbolic x coords={Criminals, Activists, States, Competitors},
			enlarge x limits=0.15,
			grid=both,
			xmajorgrids=true,
			ymajorgrids=true,
			ytick={0,20,40,60,80,100},
			legend style={
				at={(0.5,0.95)},
				anchor=north,
				legend columns=5,
				draw=none,
				font=\small,
				fill=white,
				fill opacity=0.7
			},
			nodes near coords={\pgfkeys{/pgf/fpu}\pgfmathparse{\pgfplotspointmeta}\pgfmathprintnumber{\pgfmathresult}}, 
			every node near coord/.append style={
				font=\small,
				yshift=12pt,
				anchor=center,
				rotate=90,
				text=black
			},
		]
			
			\definecolor{color5}{RGB}{160,193,90}
			\definecolor{color4}{RGB}{173,214,51}
			\definecolor{color3}{RGB}{255,217,52}
			\definecolor{color2}{RGB}{255,178,52}
			\definecolor{color1}{RGB}{255,140,90}
			
			\addplot+[ybar, fill=color1, draw=black, bar shift=-20 pt] plot coordinates {
				(Criminals, 0)
				(Activists, 9)
				(States, 0)
				(Competitors, 0)
			}; 
			
			\addplot+[ybar, fill=color2, draw=black, bar shift=-10 pt] plot coordinates {
				(Criminals, 9)
				(Activists, 27)
				(States, 20)
				(Competitors, 27)
			}; 
			
			\addplot+[ybar, fill=color3, draw=black, bar shift=0 pt] plot coordinates {
				(Criminals, 27)
				(Activists, 27)
				(States, 20)
				(Competitors, 45)
			}; 
			
			\addplot+[ybar, fill=color4, draw=black, bar shift=10 pt] plot coordinates {
				(Criminals, 36)
				(Activists, 18)
				(States, 20)
				(Competitors, 9)
			}; 
			
			\addplot+[ybar, fill=color5, draw=black, bar shift=20 pt] plot coordinates {
				(Criminals, 27)
				(Activists, 18)
				(States, 40)
				(Competitors, 18)
			}; 
			
			\legend{Irrelevant, Slightly, Moderately, Strongly, Highly Relevant}
		\end{axis}
	\end{tikzpicture}
	}
	\caption{Expert responses on the relevance of various attackers for \ac{SDV}.}
	\label{fig:RQ1_attackers}
	\Description{Questionnaire RQ1.2 results.}
\end{figure}

As shown in \cref{fig:RQ1_answers}, experts provided feedback on various security risks related to \acp{SDV}. The highest consensus was observed for Insecure \acp{API} (S1), which exhibits a high median rating, low dispersion (IQR = 1), and a TTBA of 90.9\%, highlighting significant concerns about \acp{API} vulnerabilities. Similarly, malware in third-party applications and libraries (S2) was rated 4 or 5 by 45.5\% of the experts, underlining the risk posed by untrusted software and emphasizing the need for stringent verification processes. Concerning supply chain security (S3), the experts’ feedback was divided into hardware and software supply chain risks to capture insights in both areas. For hardware supply chains, 36.4\% of experts rated risks as highly relevant, with responses showing higher dispersion (IQR = 2). Regarding software risks, such as code injection, responses were more evenly distributed (IQR = 2), indicating that while both threat categories are acknowledged, perceptions of their criticality vary. Mixed-criticality attacks (S4) were perceived as moderately concerning, with a median rating of 3 and higher dispersion (IQR = 2), 
suggesting that this issue is considered relevant but less critical than other attack surfaces. 
In contrast, \ac{OTA} updates (S5) were rated as highly critical by a majority of experts (TTBA $>$ 70\%, IQR = 1), highlighting the importance of securing update mechanisms. 
Privacy-related risks (S6) received moderate ratings with noticeable dispersion (IQR = 2), suggesting that privacy is often perceived as secondary to security concerns in \acp{SDV}. 
Finally, 
legacy
automotive attack surfaces remain critical for \acp{SDV}. In-vehicle networks (S0-1) received high ratings with strong agreement (IQR = 1), while \ac{V2X} communication (S0-2) was rated as highly critical (score 5) by 54.5\% of experts. Smart sensors (S0-3) also received high ratings with low dispersion.

As shown in \cref{fig:RQ1_attackers}, expert feedback also highlights who the potential attackers of \acp{SDV} may be. 
Criminal actors and nation-states were perceived as the most significant threats, both receiving high ratings with low dispersion, while activists are perceived as the least relevant group. Open-ended responses further identified additional attacker profiles, including insiders, security researchers, and low-sophistication adversaries (e.g., script kiddies), suggesting a broader and more heterogeneous threat landscape.

\begin{tcolorbox}[colback=gray!10, colframe=black, boxrule=0.3pt,left=2.5pt,right=2.6pt,top=1.5pt,bottom=1.5pt]
\textbf{Takeaway 2. } 
Existing attack surfaces in \acp{CV} and \acp{AV}, such as \ac{V2X} communication and in-vehicle networks, will continue to be critical in \acp{SDV}. Emerging attack surfaces such as insecure \acp{API}, third-party software libraries, and supply chain vulnerabilities will introduce additional challenges unique to \acp{SDV}. Addressing these emerging attack surfaces alongside traditional ones is key to safeguarding SDVs as they evolve.
\end{tcolorbox}

\subsection{\ac{RQ}2 - What strategies can mitigate attack surface vulnerabilities?} \label{sec:lrrq2}

As discussed in \cref{sec:lrrq1}, the attack surfaces of \acp{SDV} span from unauthorized data access to the takeover of critical vehicle functions. This section outlines strategies to mitigate vulnerabilities across these surfaces. Rather than isolated countermeasures, the proposed mitigations are framed as design guidelines for securing \acp{SDV}, differing in architectural assumptions, lifecycle adaptability, and trade-offs such as performance overhead and certification complexity. The proposed mitigations are grouped into overarching categories. These classifications are based on solutions found in the literature and feedback from industry experts gathered through the questionnaire. Similarly to the attack surface categorization in the previous section, each mitigation category is assigned a unique identifier (M1, M2, ...) to facilitate easy reference in the text and to the corresponding attack surfaces and threats.

\noindent\textbf{M1 - \acp{IDPS}:}
\acp{SDV} can employ \acp{IDPS} to mitigate attack surfaces such as malware injection via third-party applications (S2) by monitoring internal and external communications for anomalies \cite{8514157, 10588659, 10233928}. These systems detect unauthorized access and abnormal data flows, enabling early response. However, practical deployments suffer from high false positive rates, performance overhead, and attacker evasion. Improving accuracy and robustness therefore remains an open research challenge \cite{054, 10.1145/3631204.3631864}. \ac{AI}-based \acp{IDS} for \acp{IVN} apply \ac{ML} and \ac{DL} to analyze communication patterns, including the \ac{CAN} bus, and detect both known and unknown attacks \cite{10.1145/3570954}. Nevertheless, simpler models often provide stronger baselines, as learning-based approaches trained on synthetic data may overfit and fail to generalize to real-world conditions \cite{DBLP:journals/corr/abs-2010-09470}. Expert responses highlight the importance of Gateway Firewalls \cite{pese2017hardware}, rated as highly critical by 63.6\% of participants, which, when combined with \acp{IDPS}, enable traffic filtering and real-time intrusion response. From a design perspective, deploying centralized \acp{IDS} at gateways or high-performance computing (\ac{HPC}) nodes provides global system visibility but introduces potential single points of failure, while distributed \acp{IDS} at the \acp{ECU} improve fault containment at the cost of increased management complexity. \ac{AI}-based \acp{IDS} further enhance adaptability, but raise challenges in explainability and certification, especially when trained on synthetic data \cite{WALI2025104542}.

\noindent\textbf{M2 - Secure Software Development Practices:}
A key mitigation strategy is the \ac{SVSE} Lifecycle for API and supply-chain security (S0-1, S3). Applying secure coding practices throughout \ac{SDV} software development reduces vulnerabilities introduced at design and implementation time \cite{9700220}. This includes systematic testing through static and dynamic analysis, fuzzing, and penetration testing. Adopting a secure Software Development Life Cycle (SDLC) \cite{oka2021building, 9700220}, with security embedded at all stages, is essential. Such a lifecycle incorporates threat modeling and risk assessment \cite{monteuuis2018sara, hamad2020savta, lautenbach2021proposing} to identify threats (e.g., spoofing, tampering, information disclosure), supported by automated frameworks such as \ac{TARA} \cite{9700220}. Secure and standardized vehicle \acp{API} are also required \cite{icissp17, wu2024vpi}. To mitigate third-party software risks (S2), \ac{SCA} can identify vulnerabilities in external libraries and open-source components before deployment \cite{oka2021building, 076}. Expert feedback confirms the relevance of the \ac{SVSE} life cycle, with 54.5\% of respondents rating it as highly critical. In \acp{SDV}, secure software development extends beyond pre-deployment and becomes a continuous requirement, as \ac{OTA} updates, third-party integrations, and cloud services expand the attack surface over the vehicle lifecycle. Complementing \ac{SCA}, the use of Software Bills of Materials (SBOMs), such as in UPTANE \cite{uptane}, can improve supply-chain security by providing component-level transparency and supporting vulnerability disclosure and patch management for third-party software \cite{torres-arias2019intoto, moore2022scudo-whitepaper}.

\noindent\textbf{M3 - Automotive Ethernet Security:}
In our survey, Automotive Ethernet has been identified as a key enabler for \acp{SDV}. However, Automotive Ethernet (AE) can inherit the vulnerabilities of standard Ethernet \cite{10.1145/3637059}. To mitigate security risks in AE, several strategies have been proposed to address its vulnerabilities. Firewalls can control data flow between networks, blocking unauthorized access \cite{prevelakis2015policy}. \acp{IDS} monitor network activities in real-time, identifying and preventing potential threats. Network and link-layer security mechanisms such as IPsec and MACsec are essential for ensuring data confidentiality and integrity, particularly when transmitting sensitive information within the vehicle \cite{10.1145/3637059, 10479353}. Furthermore, \ac{TLS} protects point-to-point communications, mainly securing unicast transmissions. VLANs play a key role in segmenting the network, reducing the risk of lateral movement by isolating sensitive domains \cite{10.1145/3278120, 8940022}. \ac{TESLA} and \ac{S-ARP} add further protection by preventing replay attacks and \ac{ARP} poisoning \cite{8752449}. Together, these mitigation techniques can increase Automotive Ethernet security with a relative impact on network performances. 

\noindent\textbf{M4 - \ac{OTA} Defenses:}
Both the literature and questionnaire responses emphasize a multi-layered approach to \ac{OTA} security, spanning in-vehicle and backend components (\cref{sec:lrrq3}). Effective defenses include secure boot, authenticated and integrity-protected update packages, digital signatures, role-based access control, and continuous logging and auditing. Together, these mechanisms ensure that updates are delivered, verified, and installed in a controlled and traceable manner. Encryption and authentication must be enforced throughout the update lifecycle to prevent interception, replay, or manipulation. In particular, update packages should be digitally signed and verified by the vehicle prior to installation, ensuring that only software from authorized sources is accepted and reducing the risk of malicious code injection. Lightweight public-key schemes based on elliptic curves, such as those using \ac{ECC}, are well suited for resource-constrained vehicular environments and are commonly used for key exchange and signature generation \cite{012, 10217971}. 
More decentralized approaches move verification and rollback mechanisms into the vehicle, improving resilience while complicating key management and certification.

\noindent\textbf{M5 - Data Anonymization:} 
Anonymization techniques are critical for protecting data privacy in \acp{SDV} while preserving data utility. Data masking transforms sensitive attributes, such as precise GPS coordinates or driver identifiers, through pseudonymization or obfuscation. K-anonymity \cite{10.1142/S0218488502001648} ensures that each data record is indistinguishable from at least \(k\!-\!1\) others, reducing re-identification risks. In \acp{SDV}, this can be applied to driving patterns or route data by aggregating records into similarity groups, providing cohort-level indistinguishability. L-diversity and t-closeness extend k-anonymity to address attribute disclosure. L-diversity enforces diversity of sensitive values within each group, such as driver behavior or location patterns, while t-closeness further strengthens privacy by ensuring that the distribution of sensitive attributes (e.g., fuel consumption or braking patterns) closely matches that of the overall dataset \cite{4221659}.

\noindent\textbf{M6 - Privacy-preserving Techniques:} Privacy-preserving techniques, such as Differential Privacy (DP), can provide a mathematical framework for protecting individual privacy in \acp{SDV} while preserving data utility for analytics on aggregated vehicle datasets \cite{8288389}. \acp{SDV} generate high-resolution data streams from in-vehicle sensors. For instance, by implementing differential privacy, controlled noise is systematically injected into aggregate query results, ensuring that the presence or absence of any single data point has a minimal, statistically bounded impact on the output. This probabilistic noise addition obfuscates individual-level data, preserving privacy while enabling the extraction of meaningful patterns such as driving behaviors. Differential privacy enables these analyses to remain compliant with regulatory standards by providing quantifiable privacy guarantees while upholding data utility for the data processor.

\noindent\textbf{M7 - Security‑Aware Task Orchestration and Resource Allocation:} In \acp{SDV}, this enables managing security workloads and reallocating tasks across \acp{ECU} to address cyber attacks and safety issues. Given limited \acp{ECU} resources, security mechanisms must be efficient. Hypervisor-based virtualization allows applications with different safety-criticality levels to share hardware while ensuring isolation, improving security and efficiency \cite{053}. In addition, task migration and runtime orchestration can mitigate cyber attacks or safety faults by reallocating tasks across \acp{ECU}. These approaches include static methods \cite{weiss2023predictable, baik2020poster}, where precomputed mappings for expected fault or attack scenarios are stored and activated at runtime, and dynamic methods \cite{hamad2025enhancing}, where migration decisions are made on the fly based on available resources and the affected tasks. Such mechanisms are especially effective in mixed-criticality systems, where attacks or faults may originate from components with different criticality levels. They help limit the impact of compromised tasks without requiring dedicated redundant hardware or static backups for every task.

\definecolor{LightGreen}{RGB}{160,193,90}    
\definecolor{LightYellowGreen}{RGB}{173,214,51} 
\definecolor{BrightYellow}{RGB}{255,217,52}  
\definecolor{Orange}{RGB}{255,178,52}        
\definecolor{LightRed}{RGB}{255,140,90}      
\begin{figure}[!t]
\centering
\scalebox{0.85}{
\begin{tikzpicture}
  \begin{axis}[
    xbar stacked,
    width=0.7\textwidth,
    height=2.0cm,
    bar width=15pt,
    enlargelimits=0.1,
    legend style={
      at={(1.02,0.5)},
      anchor=west,
      draw=none,
      font=\small
    },
    legend columns=1,
    xlabel={Percentage of participants (\%)},
    symbolic y coords={
      Insurance,
      Trustworthiness Scores,
      Testing,
      Ethernet Security,
      SVSE Lifecycle,
      Gateway Firewalls
    },
    ytick=data,
    y tick label style={align=right, anchor=east},
    y=0.5cm,
    nodes near coords,
    point meta=explicit symbolic,
    every node near coord/.append style={
      font=\small,
      align=center,
      inner sep=1pt,
      yshift=-3pt,
      xshift=-9pt,
      text=black
    },
    grid=none,
    xmajorgrids=true,
    ymajorgrids=false,
    grid style={dashed},
    xtick={0,10,...,100},
    xmin=0, xmax=100,
  ]

  \addplot+[xbar, fill=LightRed, draw=black] plot coordinates {
    (9.1,Insurance)[9.1]
    (0,Trustworthiness Scores)[]
    (0,Testing)[]
    (0,Ethernet Security)[]
    (0,SVSE Lifecycle)[]
    (0,Gateway Firewalls)[]
  };

  \addplot+[xbar, fill=Orange, draw=black] plot coordinates {
    (18.2,Insurance)[18.2]
    (0,Trustworthiness Scores)[]
    (0,Testing)[]
    (9.1,Ethernet Security)[9.1]
    (9.1,SVSE Lifecycle)[9.1]
    (0,Gateway Firewalls)[]
  };

  \addplot+[xbar, fill=BrightYellow, draw=black] plot coordinates {
    (63.6,Insurance)[63.6]
    (9.1,Trustworthiness Scores)[9.1]
    (36.4,Testing)[36.4]
    (0,Ethernet Security)[]
    (0,SVSE Lifecycle)[]
    (0,Gateway Firewalls)[]
  };

  \addplot+[xbar, fill=LightYellowGreen, draw=black] plot coordinates {
    (0,Insurance)[]
    (36.4,Trustworthiness Scores)[36.4]
    (45.5,Testing)[45.5]
    (45.5,Ethernet Security)[45.5]
    (54.4,SVSE Lifecycle)[54.4]
    (36.4,Gateway Firewalls)[36.4]
  };

  \addplot+[xbar, fill=LightGreen, draw=black] plot coordinates {
    (9.1,Insurance)[9.1]
    (54.5,Trustworthiness Scores)[54.5]
    (18.2,Testing)[18.2]
    (45.5,Ethernet Security)[45.5]
    (36.4,SVSE Lifecycle)[36.4]
    (63.6,Gateway Firewalls)[63.6]
  };

  \legend{
    \strut Irrelevant,
    \strut Slightly,
    \strut Moderately,
    \strut Strongly,
    \strut Highly Relevant
  }

  \end{axis}
\end{tikzpicture}
}
\caption{The experts' answers to \ac{RQ}2, where a rating of 1 indicates irrelevance and 5 indicates high relevance for \ac{SDV} security.}
\label{fig:RQ2_answers}
\Description{Questionnaire RQ2 results.}
\end{figure}

\subsubsection{\textbf{Experts' Feedback}}
Beyond the previously mentioned mitigations, industry experts suggest additional measures for improvement. 
We report some of the results of the questionnaire (\cref{fig:RQ2_answers}). 

\textbf{M8 - Testing} is a key area for improvement in \acp{SDV}: the adequacy of current testing methods received a median rating of 4 with low dispersion (IQR = 1) and TTBA = 63.6\%, indicating broad (though not unanimous) endorsement. Automated testing frameworks, such as those that generate test cases from threat models (e.g., TARA \cite{10.1145/3631204.3631864}), are critical to ensure that cybersecurity measures are comprehensive and scalable. Experts emphasized the need for more efficient testing processes that can keep up with the rapid development and deployment cycles in \acp{SDV}. This includes integrating cybersecurity testing into continuous deployment pipelines, ensuring that security is continuously monitored throughout the software lifecycle \cite{10.1145/3631204.3631864, 9700220}. 

\textbf{M9 - Trustworthiness Scores} serve as a metric to evaluate overall security of automotive software. This approach involves calculating a composite score based on various security factors, such as code quality, patching frequency, and vulnerability severity. Trustworthiness scores can also help manufacturers demonstrate compliance with standards such as ISO/SAE 21434 \cite{072}. 

\textbf{M10 - Insurance} is identified as a complementary mitigation strategy, with the literature reporting mixed views on its overall effectiveness \cite{suzanne2024insurer, bondaugwinn2023autonomous, smith2024insurance}:
experts mostly converged on moderate relevance (median = 3, IQR = 1), while strong endorsement was limited (TTBA = 9.1\%). Although some respondents view insurance as a crucial countermeasure to reduce the financial impact of security breaches, others argue that it should not be seen as a substitute for strong security practices. Instead, insurance could complement robust security measures by providing a safety net for potential damages, while manufacturers focus on implementing proactive strategies to prevent breaches.

\begin{tcolorbox}[colback=gray!10, colframe=black, boxrule=0.1pt,left=2.5pt,right=2.6pt,top=0pt,bottom=0pt]
\textbf{Takeaway 3:} \textit{Security-by-design principles} should guide the development and integration of each component, with continuous monitoring to detect evolving threats. This approach requires planning by \acp{OEM} to ensure secure vehicles in the coming years.

\textbf{Takeaway 4:} \textit{Importance of a Multi-Layered Security Approach}. The interconnected nature of \acp{SDV} demands a multilayered security architecture, incorporating in-vehicle, edge, and cloud-based measures. 

\textbf{Takeaway 5:}  \textit{Necessity of Standardized Protocols and Compliance with Regulations}. The current dependence of the industry on various protocols underscores the need for standardized approaches (e.g., following ISO/SAE 21434 and UNECE WP.29 regulations). 

\textbf{Takeaway 6:} \textit{Need for Robust Supply-Chain Security}. The \ac{SDV} ecosystem's reliance on components and software from various suppliers introduces significant risks.  
A secure supply chain requires rigorous supplier screening, secure coding practices, and regular audits. 

\textbf{Takeaway 7:} \textit{Adoption of Trustworthiness Metrics and Continuous Improvement}. As \acp{SDV} continue to evolve, trustworthiness metrics, calculated based on code quality, patch history, and other factors, offer a valuable measure of system health. 

\end{tcolorbox}

\subsection{Discussion} \label{sec:discussionVulnMit}

\begin{figure}[th!]
	\centering	\includegraphics[width=0.98\linewidth]{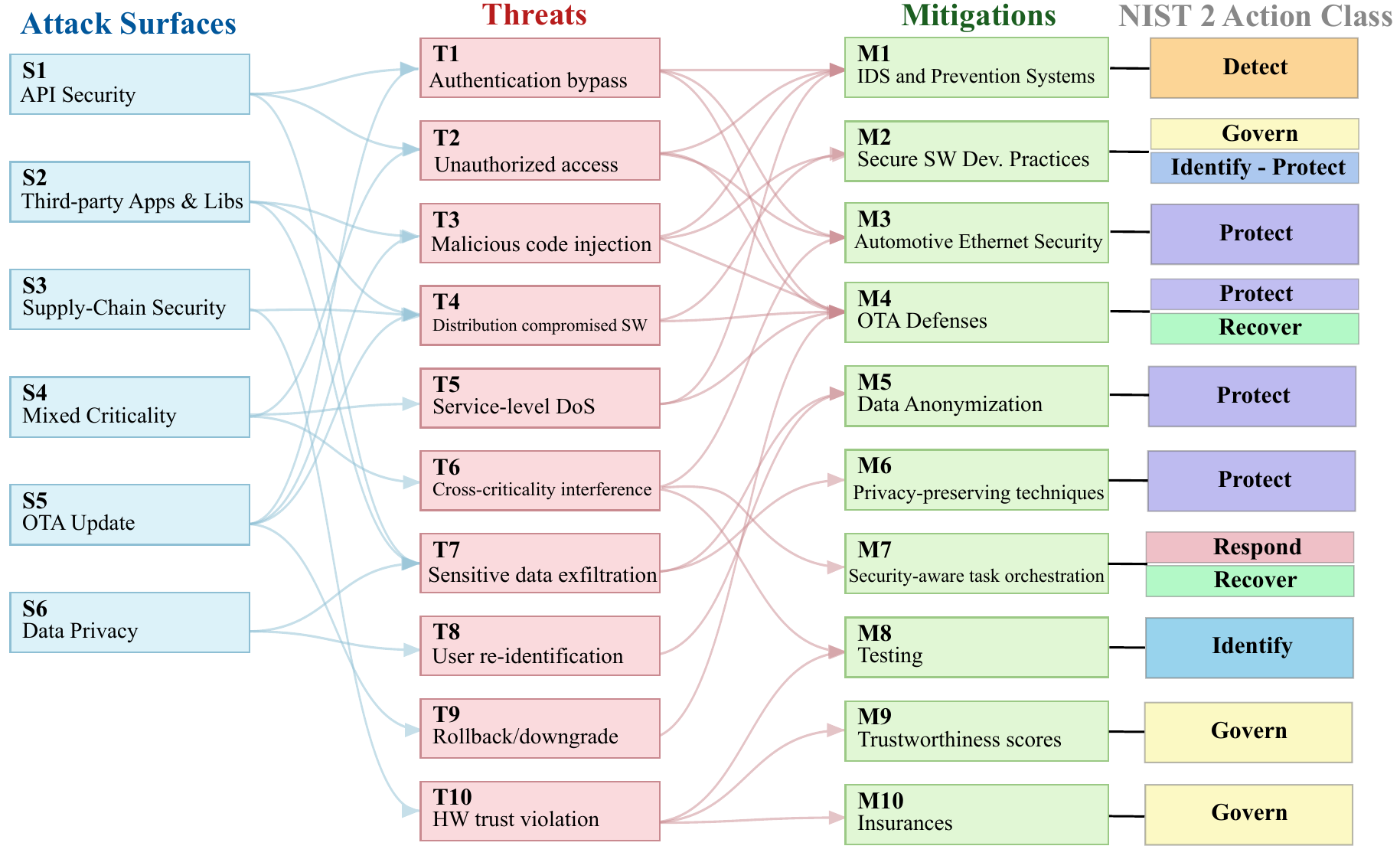}
	\caption{Conceptual mapping between \ac{SDV} attack surfaces, threat classes, and mitigation strategies.}
	\label{fig:STMschema}
	\Description{A three-column schematic showing the relationships between Software-Defined Vehicle attack surfaces, threat classes, and mitigations. Arrows connect attack surfaces to the threats they introduce, and threats to the mitigations that address them, highlighting the layered security dependencies in SDV ecosystems.}

\end{figure}

\Cref{tab:surface_threat_mitigation} and \cref{fig:STMschema} formalize a security framework grounded in the literature review and expert elicitation by organizing  heterogeneous vulnerabilities into a traceable chain from exposed interfaces (S1--S6), to adversarial objectives (T1--T10), and finally to enforceable controls (M1--M10). The visual schema captures these dependencies by aligning attack surfaces, threats, and mitigations in distinct columns and by annotating each mitigation with its corresponding \emph{NIST Cybersecurity Framework (CSF) 2.0} action class (Identify, Protect, Detect, Respond, Recover, Govern) \cite{nistCSF2024}.
This mapping contextualizes SDV-specific controls within a widely adopted, regulation-agnostic cybersecurity taxonomy. NIST CSF 2.0, released in 2024, is particularly relevant in this context because it extends the original framework with an explicit \emph{Govern} function and strengthens lifecycle-oriented risk management. 
This makes it complementary to automotive standards such as ISO/SAE~21434 and UNECE~R155/R156, which emphasize continuous risk assessment and organizational accountability. Mapping the identified mitigations onto CSF~2.0 highlights that the most security-relevant trust boundary in \acp{SDV} remains the \textit{software supply and update continuum}. The same threat classes that manifest at runtime through insecure \acp{API} (S1) and third-party components (S2), including authentication bypass and unauthorized access (T1--T2), malicious code injection and compromised distribution (T3--T4), and sensitive data leakage (T7), also reappear within the supply chain (S3) and the \ac{OTA} pipeline (S5).

As a consequence, effective mitigation cannot be addressed through isolated point solutions. Secure software development and \ac{SCA} practices (M2) primarily support the Identify and Protect functions, reducing the introduction of exploitable artifacts. Automotive Ethernet security mechanisms (M3) further strengthen the Protect function by enforcing isolation, secure communication, and traffic control across in-vehicle networks. In contrast, \ac{OTA} defenses (M4) enforce provenance, integrity, and anti-rollback guarantees (T9), effectively turning update distribution into a policy-controlled security perimeter spanning vehicle and backend services. Detection and prevention systems (M1) act as a cross-cutting Detect control that compensates for residual design and implementation flaws across S1 and S4 by monitoring service interactions and anomalous data flows. However, the framework makes explicit that detection alone cannot replace preventive guarantees where safety-critical behavior is implicated. Mixed-criticality exposure (S4) further highlights the need for evidence-driven separation. Testing and verification (M8), classified under Identify, must validate isolation properties, timing constraints, and resource budgets against cross-criticality interference (T6). Privacy-oriented mitigations (M5--M6) primarily populate the Protect function by constraining information release and bounding inference risk for T7--T8. Finally, governance mechanisms such as trustworthiness scores and insurance (M9--M10) align with the Govern function by incentivizing measurable security posture and risk transfer.

The NIST CSF~2.0 mapping also exposes a structural gap: across the surveyed literature, mitigations largely concentrate on Identify, Protect, and Govern, while explicit mechanisms addressing \emph{Respond} and \emph{Recover} remain scarce, despite being explicitly defined in the NIST CSF~2.0. This imbalance indicates that current \ac{SDV} security research prioritizes prevention and assurance over resilience and post-incident handling. Only limited efforts, such as M4 through OTA-oriented defenses and, more complete, M7 via security-aware task orchestration~\cite{hamad2025enhancing}, attempt to partially bridge this gap, which nonetheless remains largely understudied in software-driven and continuously evolving vehicular systems.

\begin{table*}[!t]
\scriptsize
\centering
\renewcommand{\arraystretch}{1.15}
\setlength{\tabcolsep}{1pt}
\caption{Integrated attack Surface--Threat--Mitigation mapping.} 
\label{tab:surface_threat_mitigation}

\resizebox{\textwidth}{!}{%
\begin{tabular}{C{0.5cm} p{2.7cm} |C{0.5cm} p{4.8cm} | C{0.5cm} p{4.5cm}}
\hline
\multicolumn{4}{c}{\textbf{Attack Surfaces and related Threats (\cref{sec:lrrq1})}} & \multicolumn{2}{c}{\textbf{Surface Mitigations (\cref{sec:lrrq2})}} \\ \hline
\textbf{SID} & \textbf{Surface} & \textbf{TID} & \textbf{Threat Name} & \textbf{MID} & \textbf{Mitigation} \\ \hline

S1 & \ac{API} Security 
& T1 & Authentication bypass 
& M1 & Intrusion Detection and Prevention Systems \\
&& T2 & Unauthorized access 
& M3 & Automotive Ethernet Security \\
&& T7 & Sensitive data exfiltration and privacy leakage 
&  M4 & OTA Defenses \\ 
&& & & M6 &  Privacy-preserving techniques\\ \hline

S2 & Third-party Apps and Lib.
& T3 & Malicious code or firmware injection 
& M1 & Intrusion Detection and Prevention Systems \\
&& T4 & Distribution of compromised software 
& M2 & Secure Software Development Practices \\
&& T7 & Sensitive data exfiltration and privacy leakage 
& M6 & Privacy-preserving techniques  \\ \hline

S3 & Supply-Chain Security 
& T4 & Distribution of compromised software 
& M2 & Secure Software Development Practices \\
&& T10 & Hardware trust violation and counterfeit
& M8 & Testing \\
&&  &  & M9 & Trustworthiness scores \\
&&  &  & M10 & Insurances \\ \hline

S4 & Mixed Criticality 
& T2 & Unauthorized access 
& M1 & Intrusion Detection and Prevention Systems \\
&& T5 & Service-level \ac{DoS} and resource exhaustion 
& M3 & Automotive Ethernet Security \\
&& T6 & Cross-criticality interference affecting safety
& M4 & OTA Defenses \\
&& & & M7 & Security-aware task orchestration \\
&& & & M8 & Testing \\ \hline

S5 & \ac{OTA} Update 
& T1 & Authentication and authorization bypass 
& M4 & OTA Defenses \\
&& T3 & Malicious code or firmware injection 
&  &  \\
&& T4 & Distribution of compromised software 
&  &  \\
&& T9 & Rollback or downgrade to vulnerable system states 
&  &  \\ \hline

S6 & Data Privacy 
& T7 & Sensitive data exfiltration and privacy leakage 
& M3 & Automotive Etherenet Security \\
&& T8 & User re-identification and inference 
& M5 & Data Anonymization \\ 
&& & & M6 &  Privacy-preserving techniques\\ \hline

\end{tabular}%
}
\end{table*}
\FloatBarrier

\section{\ac{SDV} Security Challenges}\label{sec:securityChallenges} 
In this section, several critical security challenges arising from the previous \acp{RQ} are examined. The focus is placed on \ac{OTA} systems and methodologies, as well as the complexities of managing data privacy in \acp{SDV}. These challenges are derived from an in-depth analysis of current technological trends, newly identified attack surfaces, and vulnerabilities highlighted in the literature review.

\subsection{{\ac{RQ}3} 
- What are the main security issues that \ac{OTA} updates face, including the out-of-the-vehicle environment?}\label{sec:lrrq3}

Updating vehicle firmware via \ac{OTA} has become a crucial feature, as the software-driven nature of modern vehicles require continuous updates to maintain safety and security. In \acp{SDV}, \ac{OTA} updates are a fundamental capability, enabling continuous feature delivery, bug fixes, and security patches. A major challenge in \acp{SDV} is implementing secure and efficient \ac{OTA} mechanisms that address threats originating both inside and outside the vehicle, including malicious firmware introduced through supply chains. 
From a system-level perspective, \ac{OTA} mechanisms in \acp{SDV} must be analyzed not only as communication protocols but as distributed, safety-critical update workflows spanning cloud backends, edge infrastructure, and in-vehicle components. These workflows involve multiple stakeholders, including OEMs, suppliers, and cloud service providers, introducing additional trust dependencies and coordination challenges across the SDV ecosystem. This introduces new constraints compared to traditional vehicular paradigms, including (i) coordination across heterogeneous \acp{ECU} with different criticality levels, (ii) real-time constraints that limit update latency and downtime, and (iii) the need to ensure consistency and atomicity of updates across interdependent software components. These aspects highlight that \ac{OTA} security in \acp{SDV} is not only a cryptographic problem, but also a system orchestration and lifecycle management challenge. This creates a trade-off between update flexibility and system assurance, as increased update frequency and distribution complexity can amplify both operational risks and the attack surface. Although existing solutions aim to preserve the authenticity, integrity, and confidentiality of software components, they often rely on centralized, cloud-based approaches managed by \acp{OEM}. Such solutions may inadequately address \ac{SDV}-specific challenges, including decentralized, real-time updates across multiple components with minimal operational disruption. To support software maintenance, \ac{OTA} mechanisms have therefore evolved over the last decade. However, many approaches still lack essential security and privacy protections, such as hardware-backed secure boot chains to ensure firmware integrity or rollback protection against downgrade attacks. Since the 2000s, numerous supply-chain firmware attacks have been reported, e.g., a 2024 vulnerability in a Hyundai firmware update mechanism enabled integrity violations~\cite{DBLP:journals/virology/CostantinoVM24}. Such attacks remain particularly concerning, as malicious firmware can compromise hardware integrity before deployment, making detection difficult. The following section examines \ac{OTA} mechanisms in the literature, analyzing their architectures, target vehicles, and the properties they aim to guarantee.

\subsubsection{Literature Review.}
\begin{figure}[t!]
    \centering
    \includegraphics[width=0.87\linewidth]{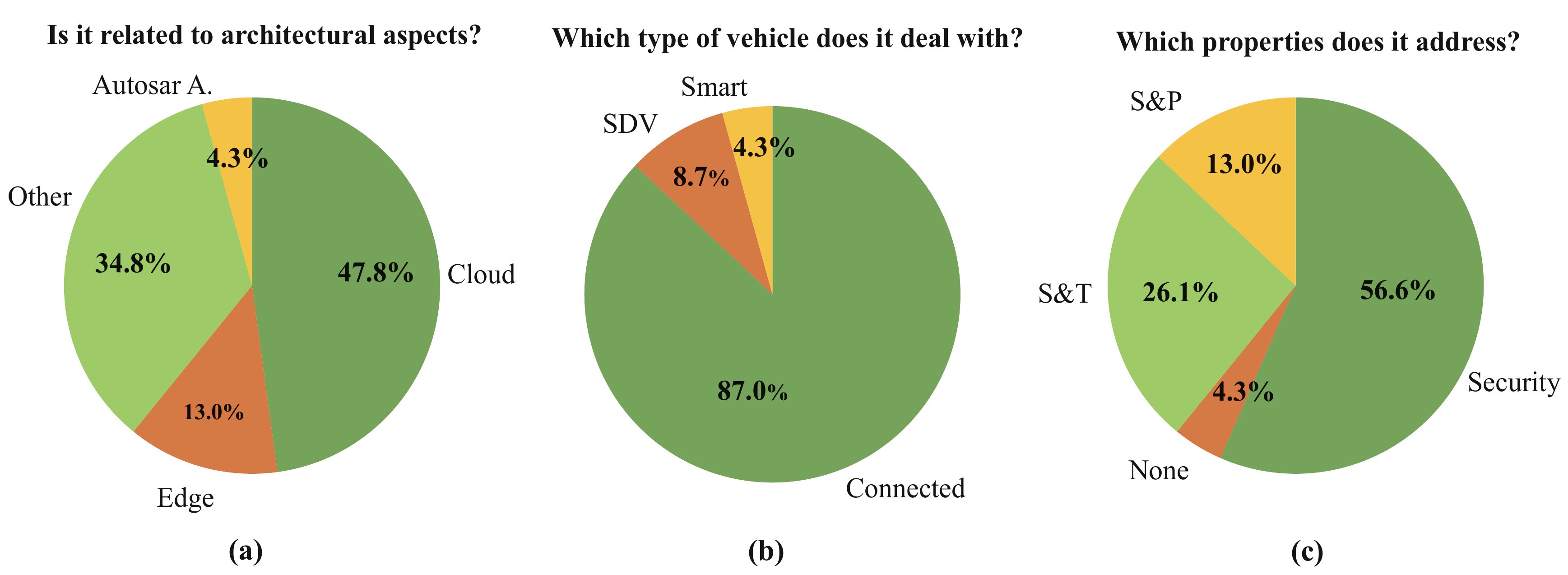}
    \caption{\ac{OTA} literature review outcomes: (a) Type of architecture, (b) Type of vehicles, and (c) Addressed properties.}
    \label{fig:ota_arch1}
    \Description{Questionnaire RQ3 results showing architectural aspects, vehicle types, and addressed properties in three pie charts.}
\end{figure}

A total of 35 articles, including research papers, white papers, surveys, and standards, were analyzed. After an initial screening, 23 papers were selected from the original set. The selection has been made by considering only papers that contribute to the advancement of the state of the art in \ac{OTA} mechanisms. 
The literature review revealed that most existing \ac{OTA} solutions focus on network security, addressing potential attackers who may try to intercept or disrupt network communications.
In particular, 87\% of the articles analyzed refer to \textit{ connected vehicles} more than to \acp{SDV} (8.7\%) denoting that network security has been considered the most crucial challenge to address (Fig. \ref{fig:ota_arch1}).

According to the literature review (Fig.~\ref{fig:ota_arch1}), 47.8\% of proposed solutions adopt a cloud-based architecture for managing \ac{OTA} processes, where software is delivered directly from \acp{OEM} to vehicles. Among these, \cite{9625372} leverages blockchain to ensure authenticity and \cite{10.1007/978-3-319-66972-4_12} proposes a blockchain-based architecture for secure \ac{OTA} updates. In 2016, UPTANE was introduced as a dedicated protocol for \ac{OTA} software updates \cite{uptane}.
Built as an extension of The Update Framework (TUF), UPTANE leverages strong cryptographic primitives to ensure the integrity, authenticity, and freshness of software updates, even under partial infrastructure compromise. It introduces a multi-role signing architecture using signed metadata, hash verification, and version constraints to mitigate threats such as machine-in-the-middle, replay, rollback, and malicious update injection. A key contribution is the separation between offline-protected signing authorities and online distribution servers. UPTANE also defines a two-tier trust model with a Primary \ac{ECU} orchestrating updates and multiple Secondary \acp{ECU} locally verifying software images before installation. This design enables defense in depth, prevents unauthorized lateral update propagation, and supports fine-grained ECU-level authorization. Since 2016, UPTANE adoption has increased, leading to extensions such as in-toto~\cite{torres-arias2019intoto} and Scudo~\cite{moore2022scudo-whitepaper}, which focus on supply-chain verification and \ac{ECU} update security, respectively.

Only 13\% of \ac{OTA} solutions for \acp{SDV} explore edge-based approaches, where updates are distributed via architectural edge nodes rather than a centralized entity. Works such as \cite{9790349} and \cite{10429931} address challenges including network availability, bandwidth, and software fragmentation. However, \cite{9790349} focuses solely on securing communication between the \ac{OEM} server and the vehicle without leveraging \ac{SDV} architecture, while \cite{10429931} does not address security. A smaller share (4.3\%) targets the AUTOSAR Adaptive Platform, which adopts a microservice architecture to support efficient software distribution. For example, \cite{10401797} proposes a secure \ac{OTA} firmware update mechanism using \ac{MQTT}, whose lightweight publish/subscribe model improves system dependability and client independence. Regarding the third subquestion (Fig.~\ref{fig:ota_arch1}), most mechanisms address security (56.5\%), while trust (26.1\%) and privacy (13\%) are considered mainly in combination with security rather than as standalone properties.

\subsubsection{\textbf{Experts' Feedback}}
\begin{figure}[t!]
	\centering
    \scalebox{0.85}{ 
	\begin{tikzpicture}
		\begin{axis}[
			ybar,
			ymin=0, ymax=100,
			width=\textwidth,
			height=0.35\textwidth,
			bar width=8pt,
			xlabel={Security elements/properties},
			ylabel={Percentage (\%)},
			xtick=data,
			symbolic x coords={Connectivity, Data Integrity, Auth.-Authoriz., Encryption, Fail-safe mech., Update mech.},
			enlarge x limits=0.15,
			grid=both,
			xmajorgrids=true,
			ymajorgrids=true,
			ytick={0,20,40,60,80,100},
			xticklabel style={font=\footnotesize},
			legend style={
				at={(0.5,0.95)},
				anchor=north,
				legend columns=5,
				draw=none,
				font=\small,
				fill=white,
				fill opacity=0.7
			},
			nodes near coords={\pgfkeys{/pgf/fpu}\pgfmathparse{\pgfplotspointmeta}\pgfmathprintnumber{\pgfmathresult}},
			every node near coord/.append style={
				font=\small,
				yshift=12pt,
				anchor=center,
				rotate=90,
				text=black
			},
		]

			\definecolor{color5}{RGB}{160,193,90}
			\definecolor{color4}{RGB}{173,214,51}
			\definecolor{color3}{RGB}{255,217,52}
			\definecolor{color2}{RGB}{255,178,52}
			\definecolor{color1}{RGB}{255,140,90}

			\addplot+[ybar, fill=color1, draw=black, bar shift=-20 pt] plot coordinates {
				(Connectivity, 9) (Data Integrity, 0) (Auth.-Authoriz., 0) (Encryption, 0) (Fail-safe mech., 0) (Update mech., 0)
			};
			\addplot+[ybar, fill=color2, draw=black, bar shift=-10 pt] plot coordinates {
				(Connectivity, 0) (Data Integrity, 9) (Auth.-Authoriz., 0) (Encryption, 9) (Fail-safe mech., 18) (Update mech., 0)
			};
			\addplot+[ybar, fill=color3, draw=black, bar shift=0 pt] plot coordinates {
				(Connectivity, 28) (Data Integrity, 18) (Auth.-Authoriz., 9) (Encryption, 27) (Fail-safe mech., 9) (Update mech., 36)
			};
			\addplot+[ybar, fill=color4, draw=black, bar shift=10 pt] plot coordinates {
				(Connectivity, 18) (Data Integrity, 9) (Auth.-Authoriz., 18) (Encryption, 37) (Fail-safe mech., 27) (Update mech., 36)
			};
			\addplot+[ybar, fill=color5, draw=black, bar shift=20 pt] plot coordinates {
				(Connectivity, 45) (Data Integrity, 64) (Auth.-Authoriz., 63) (Encryption, 27) (Fail-safe mech., 46) (Update mech., 28)
			};

			\legend{Irrelevant, Slightly, Moderately, Strongly, Highly Relevant}
		\end{axis}
	\end{tikzpicture}
    } 
	\caption{Expert responses on the significance of security elements/properties in \ac{OTA} updates.}
	\label{fig:RQ3_answers}
	\Description{Questionnaire RQ3 results.}
\end{figure}
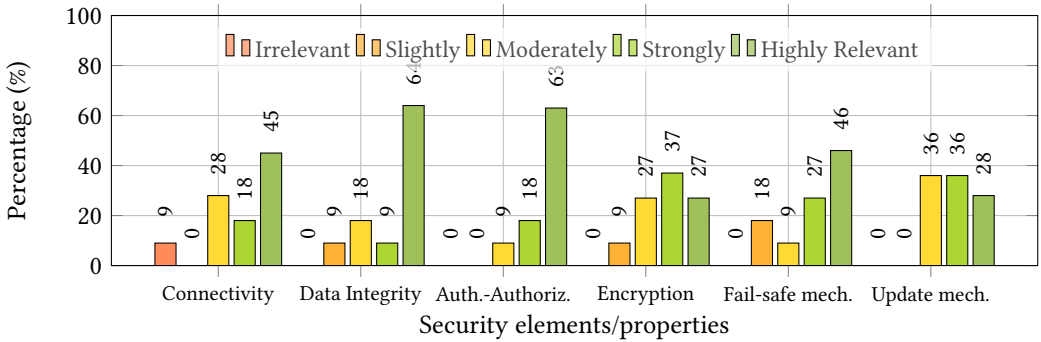

Consistent with the literature, security emerged as the most critical non-functional property of the \ac{OTA} update process (IQR = 1, TTBA $>$ 80\%). Privacy was also identified as important, though with greater dispersion, indicating less uniform prioritization. In contrast to the literature review, experts did not favor centralized cloud-based solutions; instead, Public Key Infrastructure (PKI)-based mechanisms were rated as the most effective architectural option (IQR = 1, TTBA $>$ 70\%), while centralized approaches showed lower agreement. Experts further assessed the overall criticality of \ac{OTA} updates in \acp{SDV}, rating risks such as disruption, rollback, and malicious firmware installation as highly critical (median = 5, TTBA $>$ 80\%). As shown in \cref{fig:RQ3_answers}, strong consensus was observed on key \ac{OTA} security elements, including Authentication and Authorization, Data Integrity, and Fail-Safe mechanisms (IQR = 1). Data Integrity received the highest criticality, followed by Authentication and Authorization, reinforcing the central role of security mechanisms in robust \ac{OTA} update processes. These results confirm findings from the literature that identify security as the primary non-functional requirement for \ac{OTA} updates.

\begin{tcolorbox}[colback=gray!10, colframe=black, boxrule=0.3pt,left=2.5pt,right=2.6pt,top=0pt,bottom=1.5pt]
\textbf{Takeaway 8. } 
As \acp{SDV} increasingly rely on \ac{OTA} updates for software maintenance, securing these updates is essential. Data integrity, authenticity, and secure communication channels prevent unauthorized access and ensure only verified updates are installed. Encryption, digital signatures, and access controls mitigate rollback attacks and protect software during transmission. However, strong security mechanisms can affect the performance and sustainability of \ac{OTA} processes. Effective \ac{OTA} systems must balance low overhead with robust security, with future research exploring edge computing, advanced cryptography, and distributed architectures. Importantly, \ac{OTA} security in \acp{SDV} must be addressed as a system-level problem involving architectural design, trust distribution, and lifecycle coordination, rather than as a set of isolated security mechanisms.

\end{tcolorbox}

\subsection{{\ac{RQ}4} - How do \acp{SDV} affect data collection, and what are the primary concerns related to user and vehicle privacy?} \label{sec:lrrq4}

This section reviews prior work on privacy in automotive data collection, excluding studies that jointly address security and privacy since they primarily focus on security, as commonly observed in the broader \ac{V2X} privacy literature. Vehicle data collection has been practiced for over a decade by both \acp{OEM} \cite{bmw_cardata,otonomo} and third-party actors such as Usage-Based Insurance (UBI) providers \cite{progressive,allstate,esurance,statefarm}. In both cases, only a selection of vehicle data was collected for vendor-specific purposes. These early forms of telematics data collection platforms were either (i) only available on upper-tier models of certain \acp{OEM} or (ii) as a hardware dongle that accesses publicly available sensor data through OBD-II.
From a system perspective, data privacy in \acp{SDV} differs from traditional vehicular contexts due to continuous data generation, cloud integration, and cross-domain data sharing. Privacy risks therefore emerge not only from data collection, but also from data aggregation, inference, and secondary use across distributed ecosystems, third parties, and infrastructure providers. A distinguishing privacy challenge in \acp{SDV} arises from their ability to dynamically evolve data access capabilities through \ac{OTA} updates. Unlike traditional connected vehicles, where data access pathways are largely fixed at design time (e.g., via static interfaces), \acp{SDV} enable \acp{OEM} or third-party components to introduce new software modules that can subscribe to previously inaccessible sensor data. For example, an OTA update may introduce new analytics services that subscribe to raw sensor streams (e.g., camera, location, or driver behavior data) that were not previously accessible to that component, enabling new forms of profiling or inference. This creates a form of evolving data access surface, where the set of collected data is not fixed but can expand over time, often without direct user awareness. Consequently, privacy risks are not limited to initial data collection, but extend to the continuous redefinition of what data is collected, how it is used, and by whom.

Automotive data privacy has gained increased scrutiny as vehicles have become valuable sources of personal information. Mozilla reported serious privacy violations across multiple automakers \cite{mozilla2023privacy}, finding that manufacturers such as Nissan, Volkswagen, and Toyota collect sensitive data including routes, ethnicity, weight, facial expressions, and sexual behavior. Analyses of Honda’s opt-out mechanisms identified the use of dark patterns \cite{Molla2024May,bosch2016tales,gray2018dark}. Further investigations revealed that General Motors shared driving behavior data with insurance companies via OnStar without adequate customer awareness, increasing insurance costs \cite{Hill2024Apr}. 
The emergence of platforms such as \ac{AAOS}, now adopted by several major \acp{OEM} (including Honda~\cite{hondagoogle}, General Motors~\cite{gmgoogledev}, Ford~\cite{fordgoogle}, Volvo/Polestar~\cite{volvogoogle}, and Stellantis~\cite{snappauto}), further amplifies these concerns. While \ac{AAOS} inherits Android’s permission model, its access to a broader set of in-vehicle sensors and its large-scale deployment increase the potential for privacy violations.

The surveyed literature largely addresses privacy through generic techniques and does not explicitly address two challenges that are specific to the \ac{SDV} paradigm. First, \emph{telemetry retention} is typically implicit: \acp{SDV} enable persistent, high-resolution data collection over the vehicle lifecycle (e.g., \ac{OTA} logs and continuous diagnostics), raising long-term retention and secondary-use risks beyond those considered for traditional \acp{CV}. Second, \emph{consent management} is usually assumed to be static, despite the dynamic and contextual nature of consent in \acp{SDV}, where features can be enabled or disabled via OTA and vehicles may be shared among multiple users. The limited treatment of these aspects highlights an open gap in current \ac{SDV} privacy research. The following sections outline applicable privacy regulations for \acp{SDV}, review automotive privacy attacks, and discuss defense frameworks.

\subsubsection{Regulations}
\begin{figure}[t!]
    \centering
    \scalebox{0.90}{ 
    \begin{minipage}{\textwidth}
        \centering
        \begin{subfigure}[b]{0.65\textwidth}
            \centering
            \includegraphics[width=\textwidth]{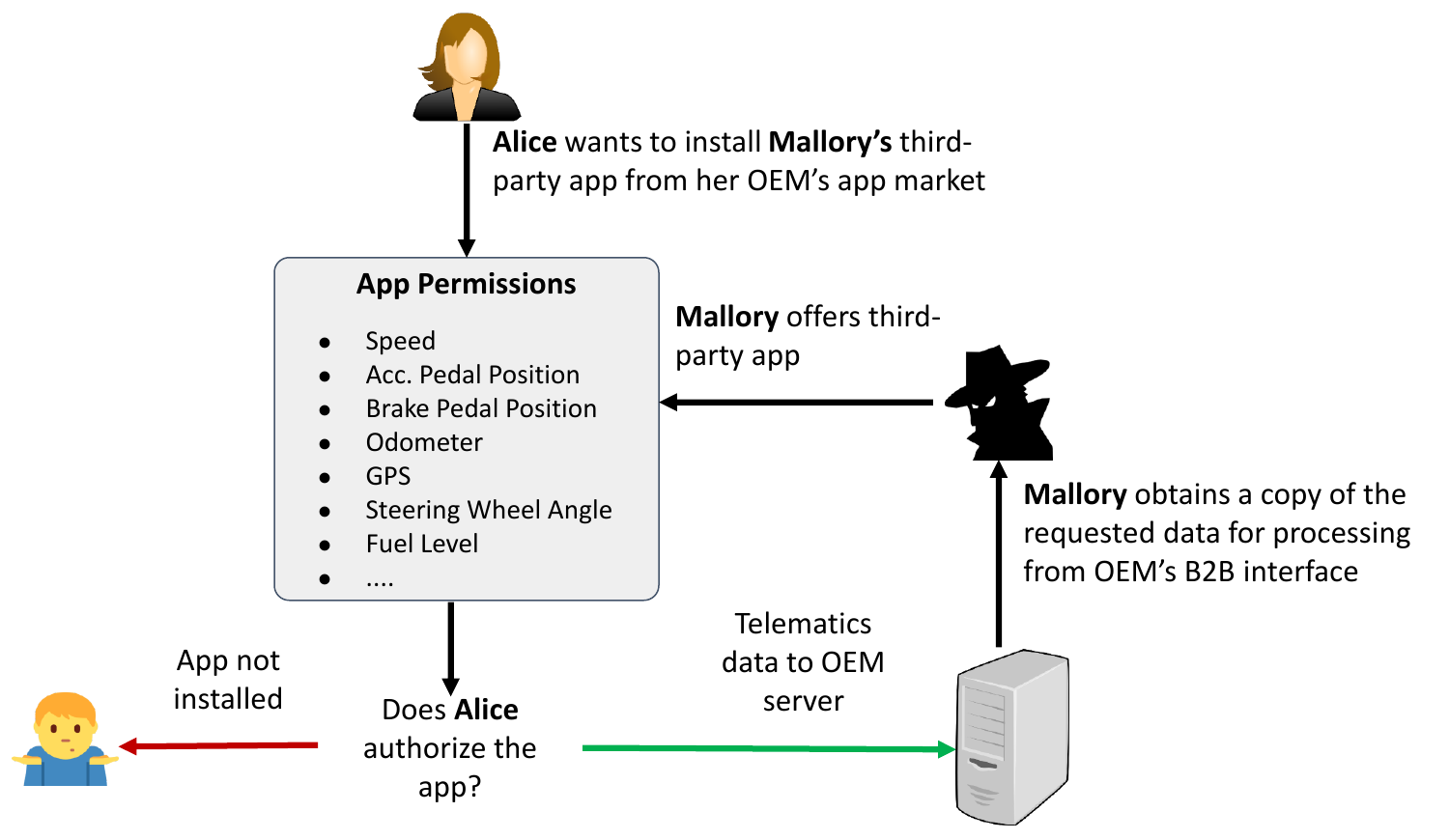}
            \caption{Data Collection in BMW CarData~\cite{bmw_cardata}}
            \label{fig:threat_model}
        \end{subfigure}
        \hfill
        \begin{subfigure}[b]{0.32\textwidth}
            \centering
            \includegraphics[width=\textwidth]{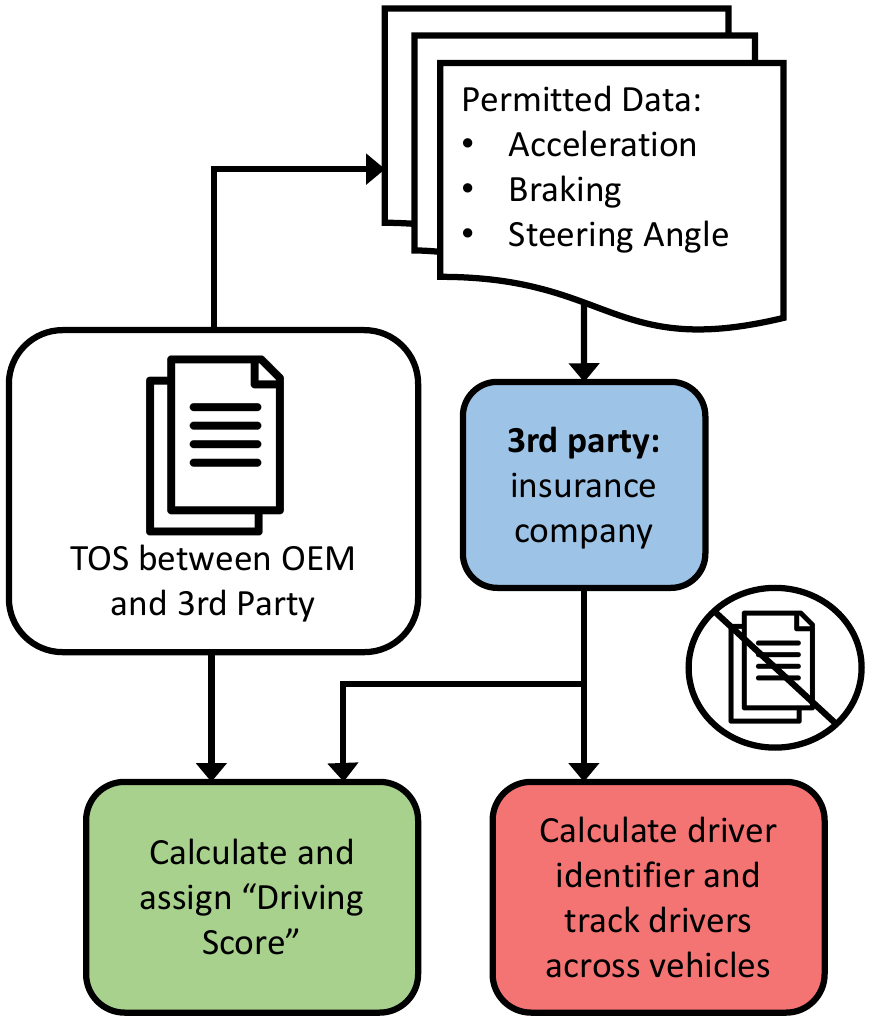}
            \caption{Data Sanitization~\cite{pese2023pricar}}
            \label{fig:example_data_sanitization}
        \end{subfigure}
    \end{minipage}
    }
    \caption{Data collection and sanitization explained.}
    \label{fig:sidebyside}
    \Description{Data Collection and Sanitization Explained.}
\end{figure}

Early automotive privacy regulation relied on voluntary guidelines from the Alliance of Automobile Manufacturers (AAM) and the 2015 Driver Privacy Act within the FAST Act in the US~\cite{pese2019survey}. In May 2018, the European Union introduced the \ac{GDPR}~\cite{GDPR} as the first comprehensive privacy framework, strengthening individual consent and data protection. Although formally applicable to EU entities, \ac{GDPR} compliance is essential for global \acp{OEM} operating across regions. In the United States, state-level laws such as the California Consumer Privacy Act (CCPA)~\cite{ccpa} and the stricter California Privacy Rights Act (CPRA)~\cite{cpra} further shape automotive data governance. The \ac{GDPR} distinguishes among \emph{data controllers}, \emph{data processors}, and \emph{data subjects}, granting strong protections to drivers as data subjects. Data controllers bear primary responsibility for determining how and why data are processed, while processors act on their behalf. \acp{OEM} are classified as data controllers due to their role in managing data shared with third-party app providers, leading to heightened compliance obligations~\cite{gdpr_automotive}. 
An evaluation of \acp{OEM} privacy policies revealed inconsistent and incomplete \ac{GDPR} implementations~\cite{pese2023pricar}. As a result, three privacy goals were identified as essential for achieving \ac{GDPR} compliance in automotive systems~\cite{pese2023pricar}:

\noindent\textbf{Data Minimization}: Current telematics platforms such as BMW CarData \cite{bmw_cardata} and \ac{AAOS} rely on permission models to limit access to specific vehicle sensors. The BMW CarData data flow, illustrated in Fig.~\ref{fig:threat_model}, is similar to that of \ac{AAOS}. However, the \ac{AAOS} permission model lacks sufficient granularity, grouping multiple vehicle properties into a small set of permissions, and assigns inadequate protection levels, reducing meaningful user consent~\cite{pese2023first}. To address this, \acp{OEM} must provide clear sensor-level explanations, and third-party apps should justify each requested permission to prevent over-privileged access.

\noindent\textbf{Data Anonymization}: When sharing the data with the third-party app provider, the privacy goal of data anonymization will eliminate any personally identifying information (PII) from the data. This can be achieved by modifying data so that it is no longer possible to identify the data subject. Pseudonymization, generalization, data masking, swapping, perturbation, and synthetic data generation are the six categories of data anonymization techniques~\cite{satori_2022}.

\noindent\textbf{Data Sanitization}: Sharing reduced or anonymized data with third parties may still violate \ac{GDPR} principles of purpose limitation and storage limitation. Once data are transferred, the \ac{OEM} loses practical control over third-party use and retention, even if constrained by the \ac{ToS}. As shown in Fig.~\ref{fig:example_data_sanitization}, an \textit{honest-but-curious} UBI provider authorized to collect braking, steering angle, and acceleration data for scoring can still perform driver fingerprinting and infer undisclosed drivers. Although illegal, such misuse cannot be technically prevented post-sharing. ISO/SAE 21434 similarly highlights that road user data are highly sensitive and easily linkable to \ac{PII}~\cite{ISO21434}.

\subsubsection{Attacks.}
A survey on inference attacks demonstrated how seemingly benign vehicular data can reveal more context than initially apparent~\cite{pese2019survey}. These inference attacks violate the GDPR Principle of Purpose Limitation and are categorized into two types: (1) \textit{Driver Fingerprinting}, (2) \textit{Location Inference} \cite{pese2019survey}. To assess the risk level of these attacks, a metric called Privacy Score (PS) was defined for 20 frequently collected vehicular sensors, indicating the potential for inferring \ac{PII}. The analysis identified location, vehicle speed, and steering wheel angle as the three most privacy-sensitive sensors.

\noindent\textit{Driver Fingerprinting.}
Driver fingerprinting aims to identify individuals operating the same vehicle using sensor data collected during trips. Prior work showed that combining 15 sensors can identify 15 drivers with 100\% accuracy~\cite{enev2016automobile}. Another study demonstrated that naturalistic driving behavior enables driver identification within minutes by training a random forest model on several hours of data~\cite{wang2017driver}. Further analysis revealed that fuel trim, brake pedal, and steering wheel data support accurate re-identification with only a few minutes of collected data using machine learning models~\cite{ezzini2018behind}. Additional work showed that drivers can be identified even before trip start, achieving high precision shortly after vehicle entry~\cite{kar2017predriveid}. More recently, driver fingerprinting was extended to raw \ac{CAN} data, reaching 97\% re-identification accuracy~\cite{gazdag2023privacy}.

\noindent\textit{Location Inference.}
Vehicle geolocation, one of the most privacy-sensitive in-vehicle data types, can be inferred from less invasive sensors such as acceleration, speed, and steering wheel angle~\cite{pese2020spy}. Early work showed that mobile \ac{IMU} data enable path inference using \ac{DFS}~\cite{dewri2013inferring}. With \ac{OBD-II} insurance dongles, Elastic Pathing enabled destination prediction from speed data~\cite{gao2014elastic}. Subsequent studies improved route identification using DFS and Hidden Markov Models~\cite{zhou2017speed,zhou2018location}. Map-matching based on distances and turning directions enabled large-scale inference without prior location knowledge~\cite{waltereit2019route}. Steering wheel angle~\cite{pese2020spy} and brake signal data~\cite{sarker2023brake} further supported route reconstruction. Recent attacks combined CAN and \ac{OBD-II} data with physical vehicle models for accurate path reconstruction~\cite{bianchi2024your}, with meter-level trajectory errors over short distances~\cite{gazdag2023privacy}.

\subsubsection{Defenses.}
Although regulation such as \ac{GDPR} or ISO/SAE 21434 can be regarded as a non-technical defense, the following subsection discusses technical defenses against aforementioned privacy attacks in academic literature.

\noindent\textit{Differential privacy} was introduced to the automotive industry as a method to mitigate privacy risks associated with vehicle data collection, providing protection for driver privacy while allowing data analysis~\cite{8288389}. This approach also addressed challenges in applying differential privacy to multidimensional time series data and proposed solutions such as personalized privacy budgets, random sampling, and event-level privacy to balance utility and privacy~\cite{8288389}. An evaluation of anonymization techniques, including smoothing, low-pass filtering, and aggregation, demonstrated their effectiveness in reducing re-identification accuracy and increasing trajectory reconstruction errors for driver fingerprinting and location inference attacks~\cite{gazdag2023privacy}. 

\noindent\textit{The PRICAR framework} was proposed to enable privacy-preserving collection and sharing of vehicle data with third parties~\cite{pese2023pricar}. PRICAR enforces data minimization, anonymization, and sanitization, with particular emphasis on sanitization techniques. It relies on a neutral entity that executes third-party code within a sandboxed environment, preventing direct data access and allowing \acp{OEM} to sanitize data before returning results. This design mitigates privacy risks and supports compliance with regulations such as \ac{GDPR}.

\noindent\textit{Another privacy-aware data access system} designed for automotive applications was developed to comply with \ac{GDPR} requirements~\cite{plappert2023privacy}. This system informs users about sensitive data flows and provides control over third-party access to personal data. Its human-machine interface (HMI) and policy framework enable users to define and manage their privacy settings while ensuring transparency and control over data usage.

\subsubsection{\textbf{Experts' Feedback}}

As shown in \cref{fig:RQ4_privacy_answers}, expert responses on privacy exhibit varying degrees of consensus. The strongest agreement concerns \textit{Opt-Out Rights for Data Collection}, with a median of 4, low dispersion (IQR = 1), and high endorsement (TTBA $\approx$ 64\%), indicating broad support for allowing drivers to opt out of data collection while retaining infotainment access. A similar pattern is observed for \textit{\ac{OEM} Responsibility for Data Sanitization} (IQR = 1, TTBA $\approx$ 64\%), reflecting the view that \acp{OEM}, as data controllers under the \ac{GDPR}, bear primary responsibility for sanitizing shared data. By contrast, opinions on the \textit{Sufficiency of Existing Privacy Mechanisms} are more divided (IQR = 2, TTBA $\approx$ 46\%), indicating disagreement on whether current infotainment privacy protections are adequate. Strong consensus is again visible for \textit{\ac{GDPR} as the Privacy Standard} (IQR = 1, TTBA $>$ 70\%), confirming its role as the dominant reference framework for privacy design in \acp{SDV}. Responses related to \textit{\ac{OEM} Control over Third-Party Apps} and \textit{\ac{OEM} Data Collection Rights} show higher dispersion (IQR = 2), highlighting ongoing debate on governance and data control. Overall, experts align on user rights and regulatory responsibility, while governance mechanisms remain more contested. 

\definecolor{LightGreen}{RGB}{160,193,90}    
\definecolor{LightYellowGreen}{RGB}{173,214,51} 
\definecolor{BrightYellow}{RGB}{255,217,52}  
\definecolor{Orange}{RGB}{255,178,52}        
\definecolor{LightRed}{RGB}{255,140,90}      

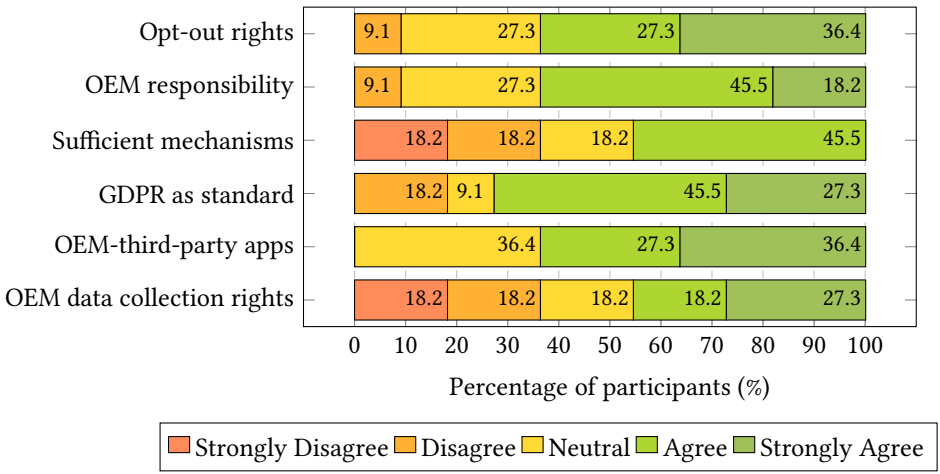
\begin{figure}[!t] 
\centering
\scalebox{0.85}{
\begin{tikzpicture}
  \begin{axis}[
    xbar stacked,
    width=0.7\textwidth,
    height=2.5cm,
    bar width=15pt,
    enlargelimits=0.1,
    legend style={
      at={(1.02,0.5)},
      anchor=west,
      draw=none,
      font=\small
    },
    xlabel={Percentage of participants (\%)},
    symbolic y coords={
      OEM data collection rights,
      OEM-third-party apps,
      GDPR as standard,
      Sufficient mechanisms,
      OEM responsibility,
      Opt-out rights
    },
    ytick=data,
    y tick label style={align=right, anchor=east},
    y=0.7cm,
    nodes near coords,
    point meta=explicit symbolic,
    every node near coord/.append style={
      font=\small,
      align=center,
      inner sep=1pt,
      yshift=-3pt,
      xshift=-9pt,
      text=black
    },
    grid=none,
    xmajorgrids=true,
    ymajorgrids=false,
    grid style={dashed},
    xtick={0,10,...,100},
    xmin=0, xmax=100,
  ]

  \addplot+[xbar, fill=LightRed, draw=black] plot coordinates {
    (18.2,OEM data collection rights)[18.2]
    (0,OEM-third-party apps)[]
    (0,GDPR as standard)[]
    (18.2,Sufficient mechanisms)[18.2]
    (0,OEM responsibility)[]
    (0,Opt-out rights)[]
  };

  \addplot+[xbar, fill=Orange, draw=black] plot coordinates {
    (18.2,OEM data collection rights)[18.2]
    (0,OEM-third-party apps)[]
    (18.2,GDPR as standard)[18.2]
    (18.2,Sufficient mechanisms)[18.2]
    (9.1,OEM responsibility)[9.1]
    (9.1,Opt-out rights)[9.1]
  };

  \addplot+[xbar, fill=BrightYellow, draw=black] plot coordinates {
    (18.2,OEM data collection rights)[18.2]
    (36.4,OEM-third-party apps)[36.4]
    (9.1,GDPR as standard)[9.1]
    (18.2,Sufficient mechanisms)[18.2]
    (27.3,OEM responsibility)[27.3]
    (27.3,Opt-out rights)[27.3]
  };

  \addplot+[xbar, fill=LightYellowGreen, draw=black] plot coordinates {
    (18.2,OEM data collection rights)[18.2]
    (27.3,OEM-third-party apps)[27.3]
    (45.5,GDPR as standard)[45.5]
    (45.5,Sufficient mechanisms)[45.5]
    (45.5,OEM responsibility)[45.5]
    (27.3,Opt-out rights)[27.3]
  };

  \addplot+[xbar, fill=LightGreen, draw=black] plot coordinates {
    (27.3,OEM data collection rights)[27.3]
    (36.4,OEM-third-party apps)[36.4]
    (27.3,GDPR as standard)[27.3]
    (0,Sufficient mechanisms)[]
    (18.2,OEM responsibility)[18.2]
    (36.4,Opt-out rights)[36.4]
  };

  \legend{
    \strut Strongly Disagree,
    \strut Disagree,
    \strut Neutral,
    \strut Agree,
    \strut Strongly Agree
  }

  \end{axis}
\end{tikzpicture}
}
\caption{Experts' answers on privacy and data collection for \ac{SDV}, where a rating of 1 represents strong disagreement and 5 indicates strong agreement.}
\label{fig:RQ4_privacy_answers}
\Description{Questionnaire RQ4 results.}
\end{figure}

\begin{tcolorbox}[colback=gray!10, colframe=black, boxrule=0.3pt, left=2.5pt, right=2.6pt, top=0pt, bottom=0pt]
\textbf{Takeaway 9. } 
Privacy concerns should be addressed with a focus on transparency, user consent and data governance. With \acp{SDV} collecting vast amounts of sensitive data, including geolocation, driver behavior, and even personal identifiers, maintaining user trust is crucial. Transparent data collection practices and clear user consent mechanisms must be integrated to address privacy concerns effectively. Providing users with control over their data can increase trust and compliance with privacy regulations while also fostering positive user engagement. A key challenge lies in balancing data utility and privacy guarantees, a trade-off that is particularly critical in \acp{SDV}, where data is central to both system operation and business models. Privacy in \acp{SDV} must account for time-evolving data access, where \ac{OTA} updates can introduce new data collection capabilities post-deployment.
\end{tcolorbox}

\section{Related work and comparison} \label{sec:rs}

Our review shows that many works retrieved using the term ``software-defined vehicle'' actually address adjacent concepts, such as software-defined vehicular networking or Internet of Vehicles, rather than \acp{SDV} as a distinct paradigm. As a result, only a limited number of surveys explicitly focus on \acp{SDV}, and existing definitions remain fragmented. In the following, we discuss the most relevant related works and position our contribution accordingly. A structured comparison is provided in \cref{tab:rwcomparison}.

\textit{SDV-focused security surveys.}
To the best of our knowledge, the first academic review explicitly addressing \acp{SDV} security is the 2023 work by Bodei \textit{et al.}~\cite{10217971}. That study investigates the transition from hardware-centric vehicles to \acp{SDV}, proposes an initial definition, and identifies emerging vulnerabilities using ISO/SAE~21434, including \ac{DoS} and jamming attacks. It highlights the limited academic attention to SDV-specific security despite strong industrial interest. Our work extends this contribution by refining the \ac{SDV} definition, expanding the analyzed attack surfaces, and incorporating \ac{OTA}, privacy, and expert-driven insights. A closely related survey is presented by Sghaier \textit{et al.}~\cite{sghaier2025advancingsecuritysoftwaredefinedvehicles}, who provide a valuable contribution by proposing a comprehensive taxonomy and threat model that map cyberattacks to \ac{SDV} architectural properties. Their work offers a detailed classification of vulnerabilities across in-vehicle, connectivity, and cloud components, helping to  structure SDV-specific security research. Our work adopts a broader systematization-of-knowledge perspective, providing an integrated treatment of security, \ac{OTA}, and privacy across the \ac{SDV} lifecycle. Moreover, it  incorporates an industry-oriented viewpoint grounded in expert elicitation and multidisciplinary academic and industrial expertise of the authors, and explicitly  proposes mitigation strategies, an aspect not addressed in previous surveys.

\textit{SDV definition and service-oriented perspectives.}
Teixeira \textit{et al.}~\cite{teixeira2024softwaredefinedvehiclesdevelopment} examine \acp{SDV} from a deterministic service-development perspective, emphasizing predictable software behavior in complex vehicles. More recently, Mate \textit{et al.}~\cite{Mate2025} bridge industrial and academic perspectives on SDVs, by discussing architectural evolution, development practices, and selected security aspects. While their work acknowledges security challenges, it does not provide a systematic threat analysis nor addresses \ac{OTA}, privacy, or lifecycle risk management in depth. In contrast, our survey integrates these dimensions within a unified framework supported by expert elicitation.

\textit{Architectural and system-level SDV discussions.}
El-Fatyany \textit{et al.}~\cite{fatyany} analyze architectural, control, and security challenges arising from software-centric vehicle designs, emphasizing the increased attack surface introduced by connectivity and centralized computation. Although aligned with several of our findings, their study does not systematically compare \acp{SDV} with other vehicle paradigms nor provides an integrated analysis of \ac{OTA} or privacy aspects.

\textit{Broader automotive software and SaaS surveys.}
Several surveys address automotive software security without explicitly targeting \acp{SDV}. For example, Huynh Le \textit{et al.}~\cite{LE201817} present a systematic survey of security and privacy in automotive software platforms, but predate the SDV paradigm and therefore do not capture its architectural and lifecycle-specific challenges. Similarly, Blanco \textit{et al.}~\cite{blanco:hal-04174527} analyze the transformation of automotive systems toward Software as a Service (SaaS), focusing on architecture, software pipelines, and runtime management. While relevant for understanding software-driven vehicle evolution, this work is not security-centric and does not address threat models or defensive mechanisms, serving instead as complementary background.

\textit{Positioning of our contribution.}
Overall, existing surveys either address vehicular cybersecurity broadly, focus on software and architectural transformation without a dedicated security lens, or propose SDV-specific taxonomies without integrating expert prioritization, privacy, and OTA-centric lifecycle analysis. This work differentiates itself by offering a holistic SDV-focused perspective that jointly considers security, \ac{OTA}, and privacy, explicitly contrasts \acp{SDV} with \acp{AV} and \acp{CV}, and incorporates expert elicitation to bridge academic analysis with industrial practice.

\begin{table}[!t]
\footnotesize
\centering
  \caption{Related work comparison.}
  \label{tab:rwcomparison}
  \renewcommand{\arraystretch}{1.05}
  \begin{tabular}{
p{2.3cm}
p{0.5cm}
C{1.1cm}
C{1.3cm}
C{1.00cm}
C{1.00cm}
C{1.00cm}
C{1.00cm}
C{1.6cm}
} \hline
   \textbf{Article} & \textbf{Year} & \textbf{Definition \ac{SDV}} & \textbf{Comparison \ac{SDV}/\ac{AV}/\ac{CV}} & \textbf{Attacks} & \textbf{Defenses} & \textbf{OTA} & \textbf{Privacy} & \textbf{Industry perspective}\\ \hline
    
    Huynh Le \textit{et al.}~\cite{LE201817} & 2018 & \ding{109} & \ding{109} & \ding{108} & \ding{108} & \ding{109} & \ding{108} & \ding{109} \\ \hline

    Bodei \textit{et al.}~\cite{10217971} & 2023 & \ding{108} & \ding{109} & \ding{108} & \ding{108} & \ding{109} & \ding{109} & \ding{109} \\ \hline

    Blanco \textit{et al.}~\cite{blanco:hal-04174527} & 2023 & \ding{108} & \ding{109} & \ding{109} & \ding{109} & \ding{108} & \ding{109} & \ding{109} \\ \hline

    Teixeira \textit{et al.}~\cite{teixeira2024softwaredefinedvehiclesdevelopment} & 2024 & \ding{108} & \ding{109} & \ding{109} & \ding{109} & \ding{109} & \ding{109} & \ding{109} \\ \hline

    El-Fatyany \textit{et al.}~\cite{fatyany} & 2024 & \ding{108} & \ding{109} & \ding{108} & \ding{108} & \ding{109} & \ding{109} & \ding{109} \\ \hline

    Mate \textit{et al.}~\cite{Mate2025} & 2025 & \ding{108} & \ding{108} & \ding{108} & \ding{109} & \ding{109} & \ding{109} & \ding{109} \\ \hline

    Sghaier \textit{et al.}~\cite{sghaier2025advancingsecuritysoftwaredefinedvehicles} & 2025 & \ding{108} & \ding{109} & \ding{108} & \ding{108} & \ding{108} & \ding{108} & \ding{109} \\ \hline

    \textbf{Our work} & 2026 & \ding{108} & \ding{108} & \ding{108} & \ding{108} & \ding{108} & \ding{108} & \ding{108} \\ \hline
     
  \end{tabular}
\end{table}

\section{Conclusion} \label{sec:conclusion}

This survey provides a comprehensive overview of cybersecurity and privacy challenges in \acp{SDV}, identifying key attack surfaces, vulnerabilities, mitigation strategies, and privacy risks. The analysis highlights the need for security frameworks capable of adapting to the software-centric evolution.

\begin{tcolorbox}[colback=gray!10, colframe=black, boxrule=0.3pt, left=2.5pt, right=2.6pt, top=0pt, bottom=0pt]
\textbf{Takeaway (Final Synthesis). }
Security and privacy in \acp{SDV} must be understood as an interconnected, system-level problem spanning attack surfaces, mitigations, \ac{OTA} mechanisms, and data governance. As highlighted in this work, specific attack surfaces, such as APIs, third-party software, supply chains, and \ac{OTA} channels, are tightly coupled with continuous software evolution and distributed architectures. These surfaces cannot be addressed in isolation: vulnerabilities in one layer (e.g., insecure APIs or supply-chain components) can propagate through \ac{OTA} updates and impact multiple vehicle functions at scale.

Mitigation strategies must therefore be coordinated across the lifecycle, combining secure software development, intrusion detection, and update verification with privacy-preserving mechanisms. Overall, ensuring trustworthy \acp{SDV} requires a holistic approach that jointly considers security and privacy across all layers of the ecosystem, from in-vehicle components to cloud services and multi-stakeholder environments.
\end{tcolorbox}

\textit{Future work}. Several research directions remain critical for strengthening \ac{SDV} security and privacy. Intrusion detection systems tailored to \acp{SDV} must evolve toward adaptive, \ac{AI}-driven approaches capable of detecting complex attacks across in-vehicle networks, \acp{API}, and \ac{OTA} processes. In parallel, supply chain security requires end-to-end frameworks that address risks arising from third-party software and hardware across the \ac{SDV} lifecycle. Given the extensive collection of sensitive data, privacy-preserving mechanisms such as differential privacy remain essential to protect \ac{PII} while enabling data-driven services. \ac{OTA} security represents another key area in which decentralized and edge-assisted approaches may improve integrity, resilience, and transparency of update processes. As \acp{API} become foundational to \acp{SDV}, interface standardization is necessary to reduce misconfigurations and exposure to exploitation. In this context, emerging \ac{AI}-based solutions, including \acp{LLM} and VLM, offer new opportunities within \ac{ITS} for security monitoring, anomaly detection, and policy reasoning, while also introducing novel attack surfaces that require careful governance, in particular in future classes of vehicles such as AIDVs \cite{10.3389/frobt.2026.1770121}. Continued alignment with standards such as ISO/SAE 21434 and UNECE WP.29 will be essential to ensure that security and privacy frameworks evolve alongside these technologies, ultimately shaping the trustworthiness of future \acp{SDV}.

\begin{acks}
We would like to thank all of the experts in the automotive industry who generously contributed their time and expertise by participating in the survey. Their valuable insights and feedback have been crucial in shaping the findings of this research. This work has been partially supported by the PNRR project Securing sOftware Platform (SOP), by the program PTR 22-24 P2.01 (Cybersecurity) and SERICS (PE00000014) under the NRRP MUR program funded by the EU, and the US National Science Foundation
Award 2345653. In addition, the work was supported by the Federal Ministry of Education and Research (BMBF) and the Free State of Bavaria under the Excellence Strategy of the Federal Government and the Länder in the context of the German-French Academy for the Industry of the Future of Institut Mines-T\'{e}l\'{e}com (IMT) and the
Technical University of Munich (TUM).
Any findings, opinions, recommendations or conclusions expressed in the article are those of the authors and do not necessarily reflect the views of the sponsor. 
\end{acks}



\bibliographystyle{ACM-Reference-Format}
\bibliography{bib/Reference}

\appendix


\end{document}